\documentclass{aa}  
\usepackage{graphicx}
\usepackage{txfonts}
\begin{document}
   \title{Manganese trends in a sample of thin and thick disk stars {\thanks{Based on observations collected at the Nordic Optical Telescope on La Palma,
Spain, and at the European Southern Observatory on La Silla, Chile,
Proposals \# 65.L-0019(B)
and 67.B-0108(B).}}\fnmsep{\thanks{The full versions of Table\,\ref{abferos.tab} 
and Table\,\ref{absofin.tab} are only available
electronically at the CDS via anonymous ftp to XXXX or via XXXX.html}}}

   \subtitle{The origin of Mn}

   \author{S. Feltzing
          \inst{1}
          \and
          M. Fohlman\inst{1}
	  \and
	  T. Bensby\inst{2}
          }

   \offprints{S. Feltzing}

   \institute{Lund Observatory, Box 43, SE-221 00 Lund, Sweden\\
              \email{sofia@astro.lu.se}
         \and
              Department of Astronomy,
University of Michigan,
Ann Arbor, MI,
USA \\
             \email{tbensby@umich.edu }
                          }

   \date{Received XXXX; accepted XXXX}

 
  \abstract
{Manganese is an iron-peak element and
  although the nucleosynthesis path that leads to its formation is
  fairly well understood, it remains unclear which objects, SN\,II
  and/or SN\,Ia, that contribute the majority of Mn to the
  interstellar medium. It also remains unclear to which extent the
  supernovae Mn yields depend on the metallicity of the progenitor
  star or not.  }
  {By using a well studied and well defined sample of 95 dwarf stars
    we aim at further constraining the formation site(s) of Mn.}
   {
We derive Mn abundances through spectral synthesis of four
\ion{Mn}{i} lines at 539.4, 549.2, 601.3, and 601.6\,nm.
Stellar parameters and data for  oxygen
are taken from Bensby et al.\,(2003, 2004, 2005).}
{When comparing our Mn abundances with O abundances for the same stars
  we find that the abundance trends in the stars with kinematics of
  the thick disk can be explained by metallicity dependent yields from
  SN\,II. We go on and combine our data for dwarf stars in the disks
  with data for dwarf and giant stars in the metal-poor thick disk and
  halo from the literature. We find that dwarf and giant stars show
  the same trends, which indicates that neither non-LTE nor
  evolutionary effects are a major concern for Mn. Furthermore, the
  [Mn/O] vs [O/H] trend in the halo is flat.  }
   {We conclude that the simplest interpretation of our data is 
that Mn is most likely produced in SN\,II and that the 
Mn yields for such SNae must be metallicity dependent. Contribution from
SN\,Ia in the metal-rich thin disk can not, however, be excluded. 
}

   \keywords{Stars: abundances -- Galaxy: abundances -- Galaxies: abundances --
(Stars:) supernovae: general 
               }

   \maketitle
%

\section{Introduction}
\label{sect:intro}

Manganese is an element in the lower iron group. It has only one
stable isotope, $^{55}$Mn, which is thought to be mainly produced in
the processes of explosive silicon burning and nuclear statistical
burning (Woosley \& Weaver 1995). Even though the nuclear path that
leads to $^{55}$Mn is fairly well understood it remains unclear which
objects that are the main contributors to the chemical enrichment for this
element and if the Mn yields from either supernovae type Ia (SN\,Ia)
or supernovae type II (SN\,II) are metallicity dependent
(e.g. McWilliam et al. 2003; Carretta et al. 2004; Woosley \& Weaver
1995).

Gratton\,(1989) investigated metal-poor stars and showed that the run
of [Mn/Fe] vs. [Fe/H] mimics that of the $\alpha$-elements but
``up-side-down'', i.e. that [Mn/Fe] is under-abundant in
low-metallicity stars but that the trend subsequently starts to
increase towards the solar value. In the study by Gratton\,(1989) this
increase starts at [Fe/H] = --1. Based on these observations they
suggested that Mn is made in SN\,Ia to a much larger degree than in
SN\,II, hence this leads to an increase at higher metallicities when
the SN\,Ia starts to contribute to the chemical enrichment of the
interstellar medium from which the subsequent generations of stars are
formed.

However, nucleosynthesis calculations (e.g. Arnett 1971; Woosley \&
Weaver 1995; Chieffi \& Limongi 2004; Limongi \& Chieffi 2005) show
that the yields from SN\,II likely are metallicity dependent. In such
a case the trends observed by e.g. Gratton\,(1989) could be
explained by higher yields from SN\,II with metal-rich progenitor
stars.

McWilliam et al.\,(2003) considered Mn abundances in three different
stellar populations: a sample of Galactic bulge K giants, giant stars
belonging to the Sagittarius stream, and a sample of solar
neighbourhood stars. The latter was taken from Nissen et al.\,(2000)
but with improved treatment of hyperfine structure (Prochaska \&
McWilliam 2000). From their comparison of these three stellar
populations McWilliam et al.\,(2003) concluded that Mn is produced
both in SN\,Ia as well as in SN\,II and the Mn yields from both types
of SNe are metallicity dependent.

Carretta et al.\,(2004) revisited the arguments of McWilliam et
al.\,(2003) and argued that by including two red giant stars in the
Sagittarius (Sgr) dwarf spheroidal galaxy, studied by Bonifacio et
al.\,(2000), the case for metallicity dependent SN\,Ia yields was
considerably weakened.

In this paper we study two local stellar samples, representative of
the thin and the thick disk, respectively, that have been shown to
have different elemental abundance trends for a number of elements
(Bensby et al.\,2003, 2005; Bensby \& Feltzing 2006).

Of particular interest for our study is that the trends of oxygen
abundances for these stars are well separated and tight (Bensby et
al.\,2004). This means, since oxygen is only produced in SN\,II, that
we have access to a ``clock'' that is independent of the SN\,Ia
time-scale. Combining this information with the new Mn abundances and
data for halo stars and stars in dwarf spheroidal galaxies (dSph) we
investigate the origin of Mn.

The paper is organized as follows. In Sect.\,\ref{sect:stars} the two
stellar samples are defined and the observations described.
Section\,\ref{sect:abun} describes the abundance analysis in detail
including an extended discussion of error sources. After that
Sect.\,\ref{sect:res} describes the resulting abundance trends and
these are discussed and interpreted in Sect.\,\ref{sect:interp}.
Finally, Sect.\,\ref{sect:sum} summarizes our findings.

\section{Description of  the stellar samples and observations}
\label{sect:stars}

\begin{table*}
\caption{Stellar parameters for the stars.  Columns 1 and 8 lists the
  Hipparcos numbers for the stars (for a cross correlation with
  e.g. HD and HR numbers see Bensby et al. 2003, 2004, 2005) Columns 2
  and 9 give the $T_{\rm eff}$, columns 3 and 10 $\log g$, and columns
  4 and 11 the [Fe/H] as derived in Bensby et al.\,(2003, 2004) and
  columns 5 and 12 the [Mn/H] abundances derived in this work Columns
  6 and 13 indicate which disk each star is assigned to. The
  assignments are the same as in Bensby et al.,(2003, 2005). 1
  indicated a thin disk star whilst a 2 indicates a thick disk
  star. Finally, columns 7 and 14 indicates which spectrograph was
  used for the observations. S = SOFIN on the NOT, F = FEROS on ESO
  1.5m (see Sect.\,\ref{sect:stars}).  }
\label{stars.tab}      
\centering          
\begin{tabular}{llll l l l llllllll}     
\hline\hline       
ID & $T_{\rm eff}$ &  $\log g$ & [Fe/H] & [Mn/H] & D. &S. &&ID & $T_{\rm eff}$ &  $\log g$ & [Fe/H] &  [Mn/H] & D. & S. \\  
& & [K] &  &  & & & && [K] &  &  &  \\
\hline                    
Sun	 & 5777 & 4.44  & +0.00  &               & 1   &F && HIP80337   & 5880 &4.49 &   +0.03 & --0.08 &1 &F\\        
Sun	 & 5777 & 4.44  & +0.00	 &               & 1   &S && HIP80686   & 6090 &4.45 &  --0.06 & --0.14 &1 &F\\     
HIP699	 & 6250 & 4.19  & --0.20 & --0.24	 & 1   &S && HIP81520   & 5680 &4.53 &  --0.48 & --0.60 &1 &F\\ 
HIP910	 & 6220 & 4.07  & --0.36 & --0.43	 & 1   &S && HIP82588   & 5470 &4.55 &  --0.02 & --0.06 &2 &F\\     
HIP2235	 & 6645 & 4.17  & --0.28 &	         & 1   &S && HIP83229   & 5770 &4.17 &  --0.57 & --0.75 &2 &F\\     
HIP2787	 & 6620 & 3.86  & --0.11 & --0.14	 & 1   &S && HIP83601   & 6167 &4.48 &   +0.09 &  +0.10 &1 &F\\     
HIP3086	 & 5840 & 4.15  & --0.11 & --0.21	 & 2   &F && HIP84551   & 6475 &3.79 &   +0.12 &  +0.12 &1 &F\\     
HIP3142	 & 6100 & 4.07  & --0.45 & --0.52	 & 1   &F && HIP84636   & 5820 &3.91 &   +0.23 &  +0.30 &1 &F\\     
HIP3185	 & 5320 & 3.78  & --0.59 & --0.80	 & 2   &F && HIP84905   & 5830 &4.06 &  --0.57 & --0.72 &2 &F\\     
HIP3497	 & 5636 & 4.30  & --0.33 & --0.46	 & 2   &F && HIP85007   & 6050 &4.37 &  --0.39 & --0.51 &1 &F\\     
HIP3704	 & 6040 & 4.30  & --0.38 & --0.47	 & 2   &F && HIP85042   & 5720 &4.40 &   +0.03 &  +0.04 &1 &F\\     
HIP3909	 & 6270 & 4.41  & --0.06 & --0.11	 & 1   &S && HIP86731   & 5840 &3.79 &   +0.25 &  +0.34 &1 &F\\     
HIP5315	 & 5030 & 3.46  & --0.42 & --0.58	 & 2   &F && HIP86796   & 5800 &4.30 &   +0.32 &  +0.36 &1 &F\\     
HIP7276	 & 5930 & 3.99  & +0.20	 &  +0.31        & 1   &F && HIP87523   & 6070 &4.07 &  --0.40 & --0.47 &1 &F\\     
HIP9085	 & 6200 & 4.25  & --0.31 & --0.42	 & 1   &F && HIP88622   & 5720 &4.35 &  --0.46 & --0.64 &2 &F\\     
HIP10306 & 6560 & 4.04  & --0.17 & --0.16	 & 1   &S && HIP88945   & 5690 &4.37 &  --0.05 & --0.21 &1 &S\\     
HIP10798 & 5350 & 4.57  & --0.47 & --0.54	 & 1   &F && HIP90485   & 6339 &4.05 &   +0.12 &  +0.15 &1 &F\\     
HIP11309 & 6210 & 3.98  & --0.32 &               & 2   &S && HIP91438   & 5580 &4.42 &  --0.24 & --0.31 &1 &F\\     
HIP12186 & 5800 & 4.04  & +0.14	 &  +0.10        & 1   &F && HIP92270   & 6370 &4.32 &  --0.06 & --0.09 &1 &S\\     
HIP12306 & 5765 & 4.20  & --0.53 & --0.67	 & 2   &S && HIP93185   & 5810 &4.40 &  --0.28 & --0.39 &1 &S\\     
HIP12611 & 5250 & 3.66  & +0.26	 &  +0.21        & 1   &F && HIP94645   & 6200 &4.35 &   +0.16 &  +0.21 &1 &F\\     
HIP12653 & 6150 & 4.37  & +0.14	 &  +0.13        & 1   &F && HIP96124   & 5590 &4.37 &  --0.20 & --0.30 &2 &F\\     
HIP14086 & 5110 & 3.51  & --0.59 & --0.76	 & 2   &F && HIP96258   & 6380 &4.25 &  --0.02 & --0.09 &1 &S\\     
HIP14954 & 6240 & 4.10  & +0.19	 &  +0.36        & 1   &F && HIP96536   & 6397 &4.06 &   +0.03 &  +0.06 &1 &F\\     
HIP16788 & 5920 & 4.24  & --0.32 & --0.42        & 2   &S && HIP98767   & 5490 &4.23 &   +0.25 &  +0.20 &2 &F\\     
HIP17147 & 5920 & 4.33  & --0.84 & --1.02	 & 2   &F && HIP98785   & 6430 &3.97 &   +0.03 &  +0.12 &1 &F\\     
HIP17378 & 5020 & 3.73  & +0.24	 &  +0.17        & 1   &F && HIP99240   & 5585 &4.26 &   +0.37 &  +0.39 &1 &F\\     
HIP18235 & 5000 & 3.13  & --0.71 & --0.94	 & 2   &S && HIP102264  & 5570 &4.37 &  --0.22 & --0.34 &1 &F\\     
HIP18833 & 6370 & 3.99  & --0.52 & --0.54 	 & 1   &S && HIP103458  & 5780 &4.30 &  --0.65 & --0.86 &2 &F\\     
HIP20242 & 5650 & 3.94  & --0.26 & --0.41        & 2   &S && HIP103682  & 5940 &4.26 &   +0.27 &  +0.34 &1 &F\\     
HIP21832 & 5570 & 4.27  & --0.61 & --0.78	 & 2   &S && HIP105858  & 6067 &4.27 &  --0.73 & --0.86 &1 &F\\     
HIP22263 & 5850 & 4.50  & +0.05	 & --0.04        & 1   &F && HIP107975  & 6460 &4.06 &  --0.53 &1 &S\\     
HIP22325 & 6250 & 3.91  & +0.06	 &  +0.08        & 1   &F && HIP108736  & 5890 &4.24 &  --0.38 & --0.54 &2 &F\\     
HIP23555 & 6300 & 4.29  & +0.13	 &  +0.16        & 1   &F && HIP109378  & 5500 &4.30 &   +0.22 &  +0.20 &1 &F\\     
HIP23941 & 6427 & 4.04  & --0.30 & --0.34	 & 1   &F && HIP109450  & 5830 &4.18 &  --0.13 & --0.17 &2 &F\\     
HIP24829 & 6360 & 3.93  & +0.06	 &  +0.02        & 1   &F && HIP109821  & 5800 &4.29 &  --0.08 & --0.12 &2 &F\\     
HIP26828 & 6410 & 4.15  & --0.37 & --0.48        & 2   &S && HIP110341  & 6500 &4.29 &  --0.17 & --0.16 &1 &F\\     
HIP29271 & 5550 & 4.38  & +0.10	 &  +0.10        & 1   &F && HIP110512  & 5770 &4.15 &  --0.30 & --0.45 & 2 &F\\     
HIP30480 & 5890 & 4.15  & +0.19	 &  +0.25        & 1   &F && HIP112151  & 5035 &3.43 &  --0.42 & --0.54 &2 &S\\     
HIP30503 & 5820 & 4.37  & +0.04	 & --0.01        & 1   &F && HIP113137  & 5800 &4.10 &   +0.22 &  +0.23 &1 &F\\     
HIP36874 & 5730 & 4.22  & --0.07 & --0.16        & 2   &S && HIP113174  & 6870 &4.14 &  --0.11 &1 &S\\     
HIP37789 & 5900 & 4.09  & --0.67 & --0.82        & 2   &S && HIP113357  & 5789 &4.34 &   +0.20 &  +0.22 &1 &F\\     
HIP40613 & 5740 & 4.11  & --0.63 & --0.86	 & 2   &S && HIP113421  & 5620 &4.29 &   +0.35 &  +0.41 &1 &F\\     
HIP44075 & 5875 & 4.10  & --0.91 & --1-14	 & 2   &S && HIP116421  & 5700 &4.15 &  --0.45 & --0.62 &2 &S\\     
HIP44860 & 5690 & 4.19  & --0.45 & --0.65	 & 2   &S && HIP116740  & 5740 &3.96 &   +0.05 & --0.09 &2 &S\\     
HIP72673 & 6350 & 4.18  & --0.62 & --0.67	 & 1   &F && HIP117880  & 6100 &4.21 &   +0.12 &  +0.14 &1 &F\\     
HIP75181 & 5650 & 4.30  & --0.34 & --0.50	 & 2   &F && HIP118010  & 5795 &4.17 &  --0.07 & --0.13 &2 &S\\     
HIP78955   & 5880 &4.22 &  +0.33 &  +0.36        &1    &F && HIP118115  & 5800 &4.30 &  --0.01 & --0.11 &2 &F\\
HIP79137   & 4900 &3.62 &  +0.30 &  +0.44        &2    &F &&  \\
\hline                  
\end{tabular}
\end{table*}

The stars were selected purely based on their kinematic
properties. Briefly, we assume that the velocity components for the
halo, thick disk, and thin disk all have Gaussian
distributions. Allowing for the different asymmetric drifts of the
three populations we calculate the probability for each star that it
belongs to the halo, the thin disk, and the thick disk. For the thick
disk sample we then selected those stars that were much more likely to
be thick than thin disk stars and vice versa. In this way we selected
two fairly extreme kinematic samples of disk stars.  The full
selection procedure is detailed in Bensby et al.\,(2003, 2005). 

The stars were observed during four observing runs. Two using the
FEROS spectrograph on the ESO 1.5m telescope (in Sept. 2000 and
Sept. 2001) and two runs using the SOFIN spectrograph on the NOT
telescope (Aug. 2002 and Oct./Nov. 2002). The spectra from FEROS have
a resolution of R$\sim 48 000$ and those from SOFIN have R$\sim 80
000$. For details about data reduction see Bensby et al.\,(2003, 2005).

At each observing run we also obtained spectra of scattered solar
light.  At the FEROS runs we obtained an integrated solar spectrum by
observing the sky in the afternoon.  In the SOFIN runs we obtained a
solar spectrum by observing the Moon during the night.

Table\,\ref{stars.tab} lists the stars used in this study, their
fundamental parameters, and which spectrograph that was used to
obtain the stellar spectra.

\section{Abundance analysis}
\label{sect:abun}
\begin{figure}
\centering
\includegraphics[angle=-90,width=8cm]{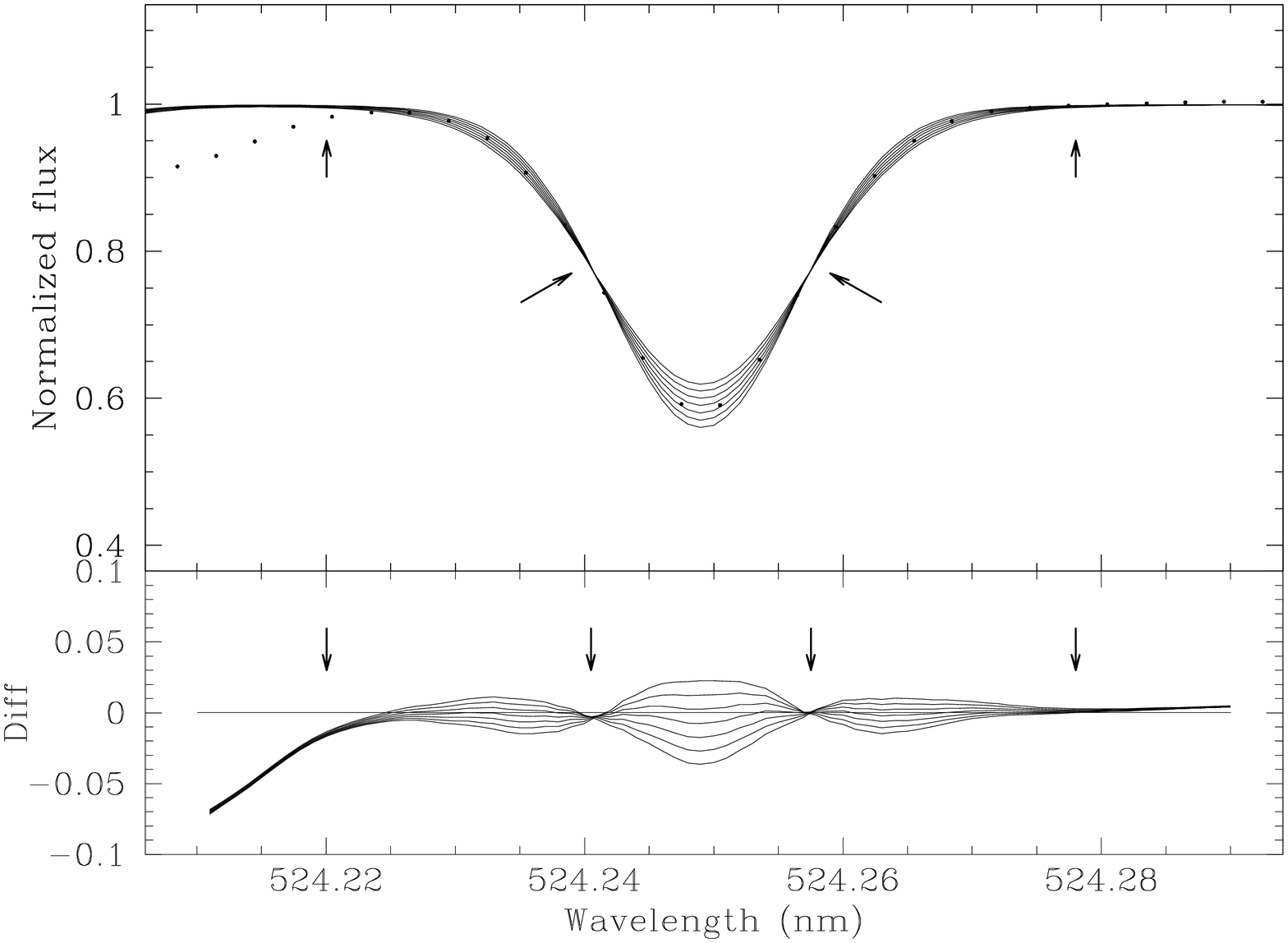}
\includegraphics[angle=-90,width=5cm]{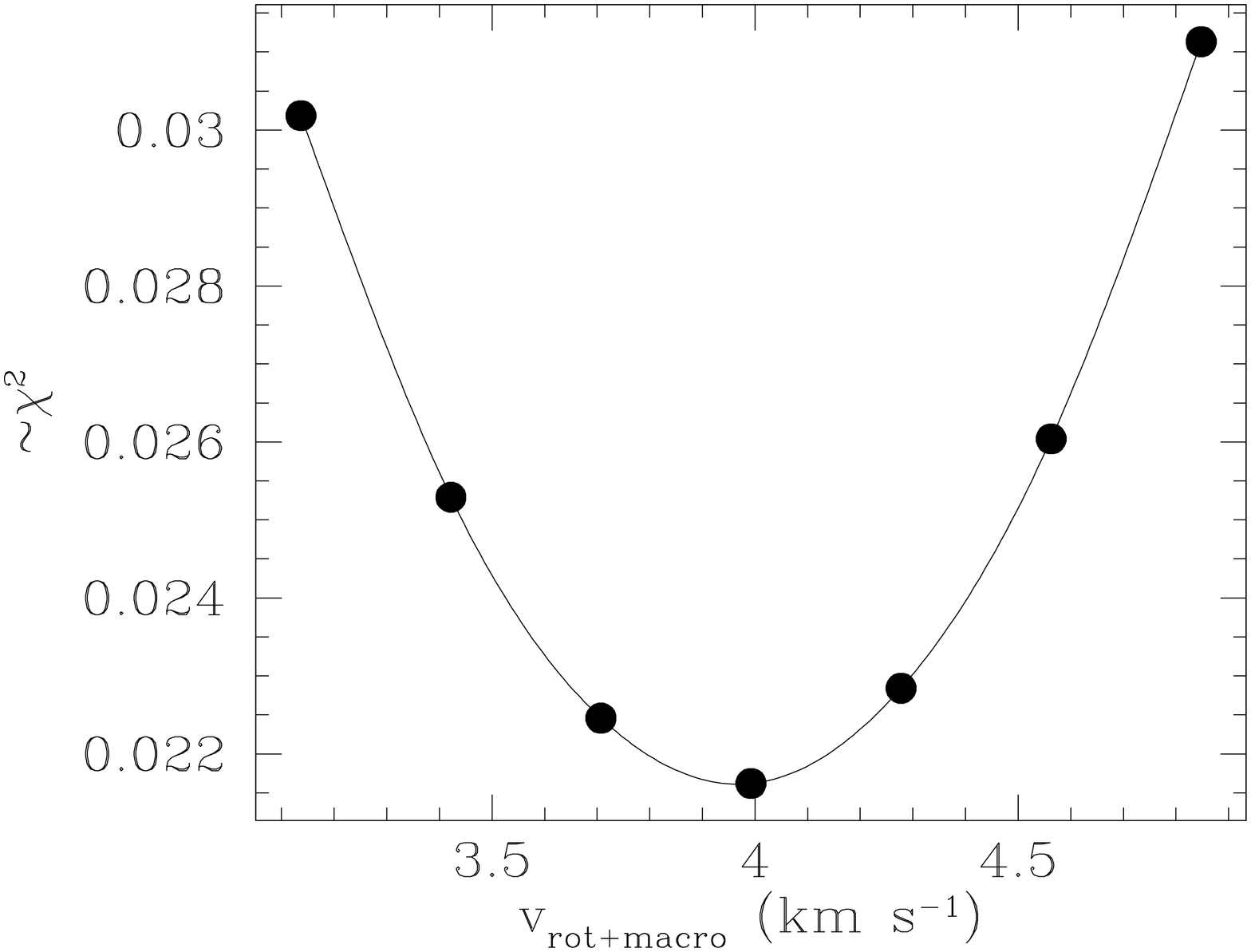}
\caption{{\bf Upper panel} Shows the observed spectrum, $\bullet$, of
  an Fe\,{\sc i} line used to derive the combined effects of rotation
  and macroturbulence.  Overlaid on the observed spectrum are the
  seven synthetic spectra. The bottom panel shows the difference
  between the observed and synthetic spectra. The synthetic spectra
  are in steps of 0.3 km\,s$^{-1}$. {\bf Lower panel} Shows the
  corresponding un-normalised $\chi^2$. }
\label{vmacrorot.fig}
\end{figure}

\begin{figure}
\centering
\includegraphics[angle=-90,width=8cm]{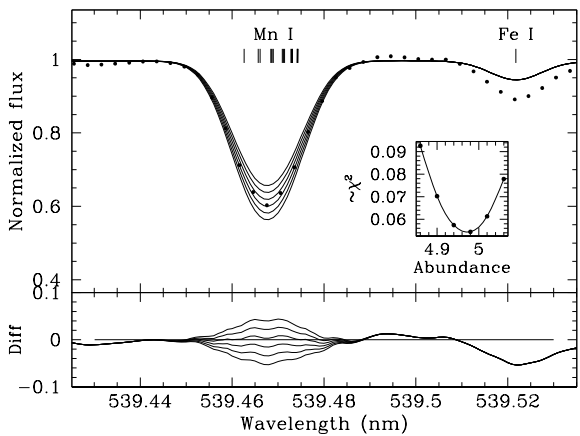}
\caption{Demonstration of the determination of the Mn abundance using
  line synthesis. Here for the Mn\,{\sc i} line at 539.4\,nm in
  HIP\,91438. The star has [Fe/H] $=-0.24$ and the spectrum a
  S/N$\sim$200, which is typical for the stars in our sample. {\bf
    Upper panel} Shows the observed spectrum, $\bullet$, and six
  synthetic spectra corresponding to six different Mn abundances,
  solid lines, in steps of 0.2 dex in the Mn abundance. The hfs for
  this line is indicated at the top with short vertical dashes, as is
  the Fe\,{\sc i} line to the right.  {\bf Lower panel} Shows the
  differences between the observed and synthetic spectra. The final Mn
  abundance can be estimated from the un-normalised $\chi ^2$, shown
  in the inset.  }
\label{ab.fig}
\end{figure}

We have performed a standard Local Thermodynamic Equilibrium (LTE)
analysis to derive chemical abundances with the help of synthetic
spectra and one-dimensional, plane-parallel model atmospheres. We have used the MARCS
model atmosphere code (Gustafsson et al. 1974; Edvardsson et al. 1993;
Asplund et al. 1997) to generate the stellar atmospheres used in the
abundance analyses.

Mn is an odd-Z element and hence its lines are subject to hyperfine
splitting (hfs) and at least for strong lines this effect must be
taken into account. As our stars span a large range in [Fe/H] and as
we know from previous studies (e.g. Gratton 1989) that Mn abundances
increases as [Fe/H] increases the Mn\,{\sc i} lines that are weak at
low [Fe/H] will be rather strong at high [Fe/H]. Hence, in order to
derive reliable abundance trends for our two samples of stars we must
include the hfs in the analysis of our spectra.

\subsection{Stellar parameters}

Stellar parameters are taken from Bensby et al.\,(2003 \& 2005).  The
effective temperatures ($T_{\rm eff}$) are based on excitation
equilibrium whilst the $\log g$ values are derived using Hipparcos
parallaxes. Full details of the iterative procedure to tune the
stellar parameters can be found in Bensby et al.\,(2003).

\subsection{Selection of lines and atomic data}
\label{sect:sellines}
\begin{table}
\caption{Selected Mn\,{\sc i} lines used in the abundance analysis
  (see Sect.\,\ref{sect:sellines} and \,\ref{linefitting.tab}). In the
  first column we give the wavelength, in the second the excitation
  energy of the lower level in the atom involved in the line
  formation, in column three the $\log gf$-values from Prochaska et
  al.\,(2000), column four lists the multiplet the line belongs to,
  and column five lists the Fe\,{\sc i} lines used to infer the
  combined macro turbulence and rotation, $v_{\rm r+m}$
  (Sect\,\ref{linefitting.tab} and Fig.\,\ref{vmacrorot.fig}).  }
\label{mnlines.tab}      
\centering          
\begin{tabular}{l l r llllllll}   
\hline\hline
&$\lambda_{\rm Mn\,I}$& $\chi_{\rm low}$ & $\log gf$ & Multiplet& $\lambda_{\rm Fe\,I}$\\ 
  &[nm]     & eV &  & & [nm] \\
\hline                    
&539.4 & 0.00 & --3.503   & 1  &  524.2, 537.9, 539.8 \\
&543.2 & 0.00 & --3.795   & 1  &  524.2, 537.9, 539.8\\
&601.3 & 3.07 & --0.251   & 27 &  606.5, 682.8\\
&601.6 & 3.07 &  +0.034   & 27 &  606.5, 682.8\\
\hline                  
\end{tabular}
\end{table}

We selected Mn\,{\sc i} lines for our analysis based on two criteria:
the lines should be easily identified and analyzed in our FEROS
spectra and they should have data for the hfs readily available in the
literature. Based on this we selected 5 lines from the list of
Prochaska et al.\,(2000). Eventually one line (at 602.1\,nm) was
discarded (see Sect.\,\ref{sect:errors} in paragraph ``Erroneous hyper
fine structure (hfs)'').  The remaining four lines are listed in
Table\,\ref{mnlines.tab}.

Atomic data for other lines in the vicinity of the Mn\,{\sc i} lines
that are needed for the line-synthesis were extracted from the Vienna
Atomic Line Database (VALD) (Kupka et al. 1999; Ryabchikova et
al. 1999; Piskunov et al. 1995).

The $\log gf$-values used in our study are the same as in Prochaska et
al.\,(2000), i.e. from Booth et al.\,(1984). These $\log gf$-values
have been experimentally determined and have a relative accuracy (as
quoted by Booth et al.\,1984) of better than 2\,\%. Furthermore, Booth
et al.\,(1984) estimate that the absolute scale for their values has
an accuracy of about 7\,\% for the ground state lines and 3\,\% for
the excited state lines.  Although more recent measurements have been
done for some Mn\,{\sc i} lines that are observable in stellar spectra
no laboratory work has been done on the lines used in this study. We
thus have no possibility for an independent, empirical assessment of
the accuracy of the $\log gf$-values.

We have, however, made comparisons of the solar abundances derived
from the lines from multiplet 1 and 27 separately. The results are
given in Eqs.\,(\ref{eq1})--(\ref{eq3}).  For the lines in multiplet 1
the differences are small, but for the lines in multiplet 27 we see a
real offset (Eq.\,\ref{eq2}).  To first order we attribute this to an
erroneous $\log gf$-value (see discussion in Sect.\,\ref{prov},
paragraph ``Absolute abundances'').

\begin{eqnarray}
{\rm FEROS: } ~ \Delta \varepsilon ({\rm Mn}) &=& +0.002 \pm 0.047 ~~~~~ \lambda:{539.4 - 543.2} \label{eq1} \\
{\rm FEROS: } ~ \Delta \varepsilon ({\rm Mn}) &=& -0.133 \pm 0.057  ~~~~~ \lambda:{601.3-601.6} \label{eq2}\\
{\rm SOFIN: } ~ \Delta \varepsilon ({\rm Mn}) &=& +0.026\pm 0.035  ~~~~~ \lambda:{539.4-543.2} \label{eq3}
\end{eqnarray}

An erroneous $\log gf$-value would affect the whole
data set and essentially provide a constant offset for all
stars for that line.
As we are doing a 
strictly  differential study with respect to the Sun the exact values 
for the $\log gf$-values is of less concern here, but clearly modern measurement
for these lines that are readily analyzed in spectra of stars with high or moderate
metallicity would be useful.

Our calculations of the stellar spectra take into account the
broadening of the stellar lines caused by collisions with hydrogen
atoms (van der Waals damping). Whenever possible, we use the data from
Barklem et al.\,(2000) for these calculations. When data are not
available from Barklem et al.\,(2000), we apply an enhancement
factor to the classical Uns\"old approximation of the van der Waals
damping, which for most elements were set to 2.5, following M\"ackle
et al. (1975). For Fe\,{\sc i} we take the correction terms from Simmons
\& Blackwell\,(1982), but for Fe\,{\sc i} lines with a lower excitation
potential greater than 2.6 eV we follow Chen et al.\,(2000) and adopt
a value of 1.4. For Fe\,{\sc ii} we adopt a constant value of 2.5
(Holweger et al.\,1990).

Two of our Mn\,{\sc i} lines have data in Barklem et al.\,(2000); the
lines at 543.2 and 539.4 nm. Experimentation shows, however, that it
does not matter for the abundance analysis which values (i.e. the
classical Uns\"old approximation or Barklem et al. data) we use for
the synthesis as these two lines are, in the Sun, completely
insensitive to any reasonable (and unreasonable) changes of the
enhancement factor or using the Barklem et al.\,(2000) data.

For the two lines in the red the situation is different. They are both
very sensitive to changes in the enhancement factor. However, these
lines are not included in the calculations by Barklem et
al.\,(2000). We can therefore only conclude that the lines are
sensitive to changes in the damping constant. However, we feel that
the value we use does give reasonable results in that the $\sigma$
(i.e. the line-to-line scatter) does for example not increase as
[Fe/H] increases (see Sect.\,\ref{sect:errors} paragraph Line-to-line
scatter).

\subsection{Fitting synthetic spectra}
\label{linefitting.tab}

\subsubsection{Broadening from rotation and macro turbulence}
\label{rotmacro.sect}

In hot stars the rotational broadening often dominates  over the
macro turbulence but in cool stars the two broadening effects  are
comparable in size (Gray 1992). Therefore the two effects can be
modeled simultaneously. To determine the rotational velocity and the
macro turbulence for the stars un-blended Fe-lines in the vicinity of
our  Mn\,{\sc i} lines were synthesized.  Data for the Fe\,{\sc i}
lines and adjacent lines  were extracted from the VALD database
(compare Sect.\,\ref{sect:sellines}).  Wavelengths for these
 Fe\,{\sc i} lines are listed in Table\,\ref{mnlines.tab}.

 The synthetic spectra were first convolved with a Gaussian profile in
 order to imitate the instrumental broadening and secondly by a
 radial-tangential (rad-tan) profile to simulate the combined effects
 from macro turbulence and rotation (Gray 1992). Before the combined
 rotational and macro turbulence velocity ($v_{\rm r+m}$) could be
 determined the $\log gf$-values for the Fe\,{\sc i} lines had to be
 adjusted in order to achieve a good fit to the observed spectrum.
 The fit was deemed good when the intersection points between the
 observed spectra and the synthetic spectra lied at zero and the
 difference between synthetic spectra and observed spectra, where the
 wings reach the continuum, also were at zero (see
 Fig.\,\ref{vmacrorot.fig} and compare e.g. Bensby et al.\,2004).

 To derive the $v_{\rm r+m}$ we calculated seven synthetic spectra
 with different $v_{\rm r+m}$ (in steps of $0.3~~ {\rm km}~~{\rm
   s}^{-1}$) for each Fe\,{\sc i} line. The synthetic spectra were
 then compared with the observed spectrum and an un-normalised ($\sim
 \chi ^2$) was calculated for each comparison (i.e.  observed -
 synthetic) and the $v_{\rm r+m}$ was read off as the value that
 minimizes $\sim \chi^2$.  The procedure is illustrated in
 Fig.\,\ref{vmacrorot.fig}.

\subsubsection{Derivation of stellar abundances}
\label{sect:derab}

\begin{figure*}
\centering
\includegraphics[angle=-90,width=7cm]{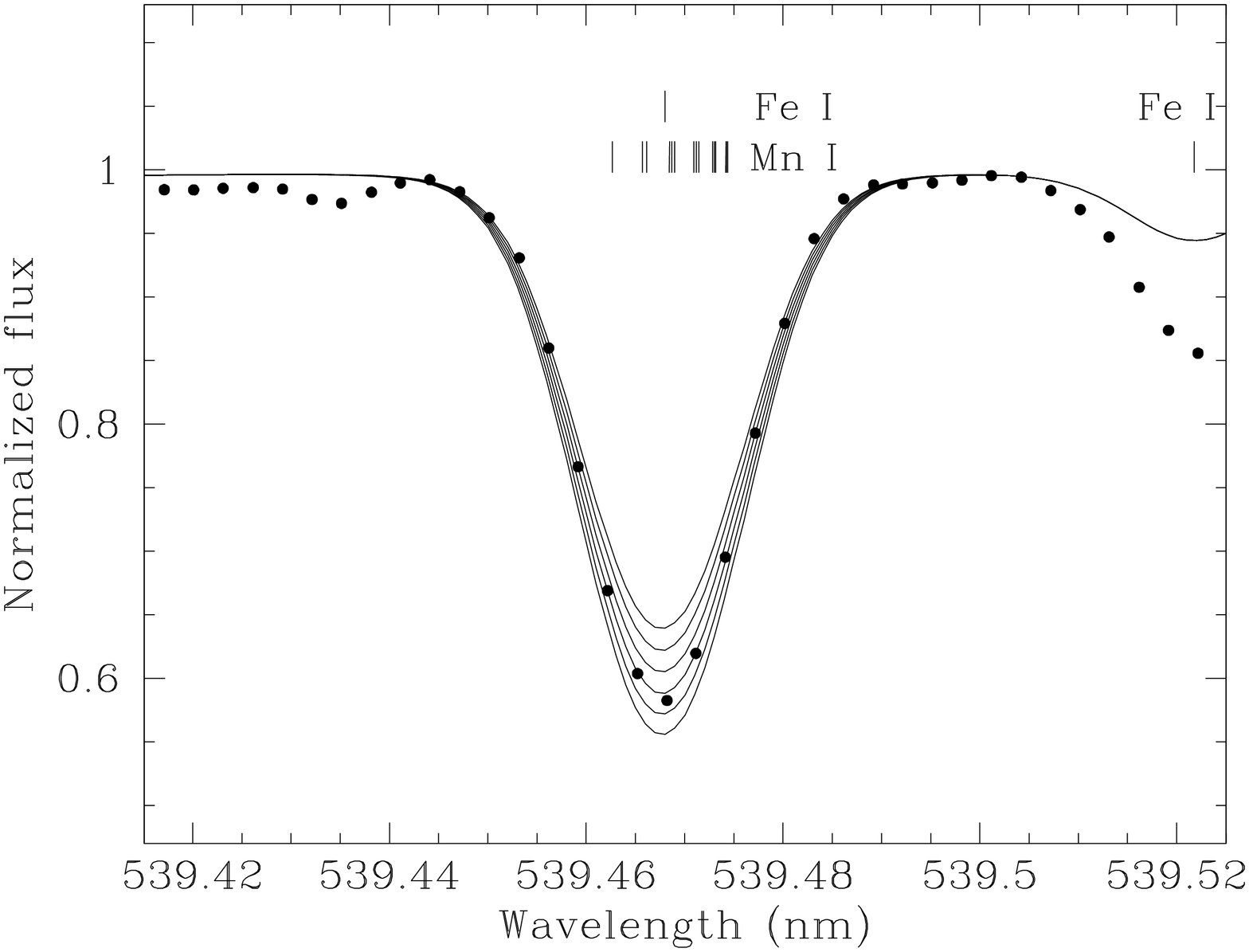}
\includegraphics[angle=-90,width=7cm]{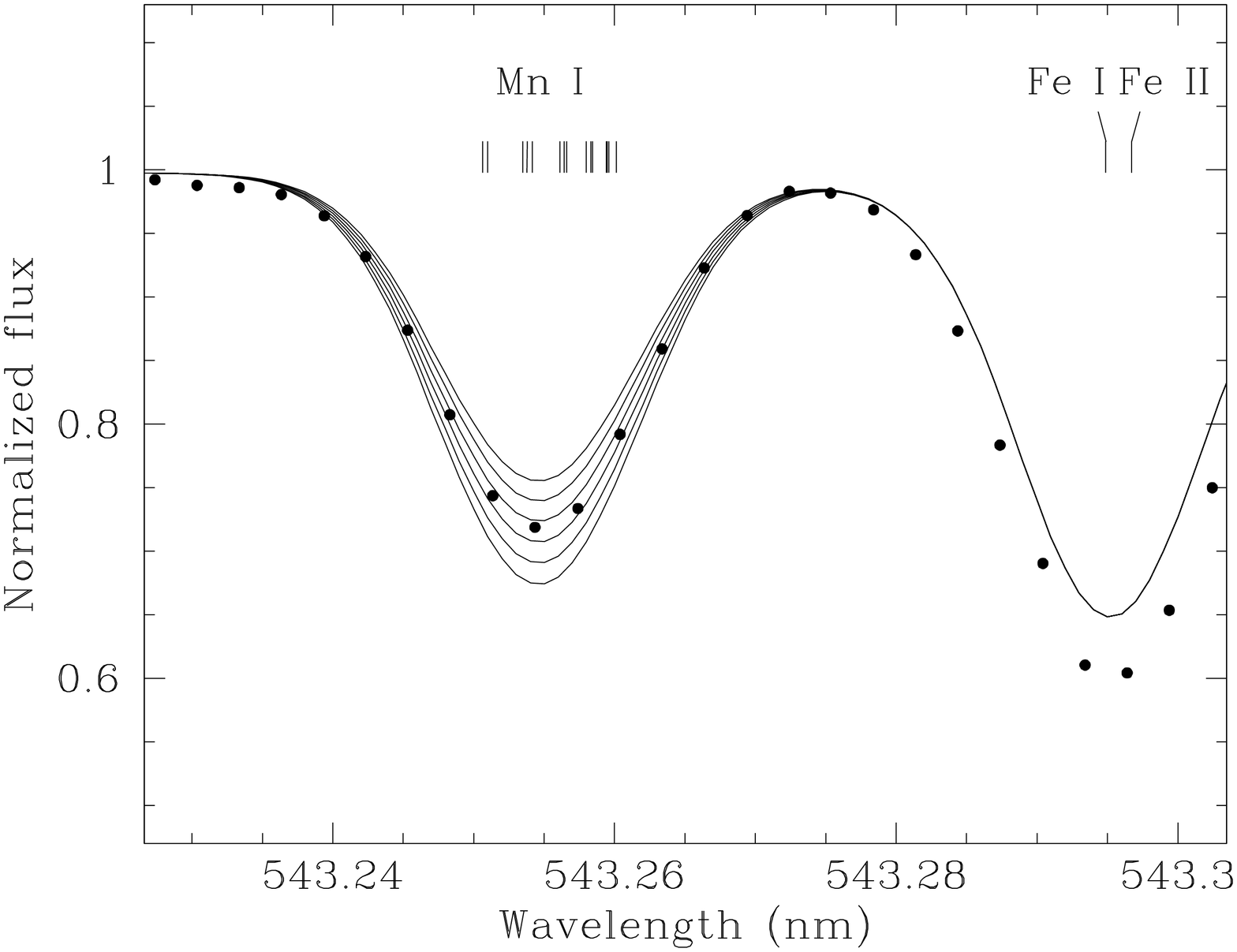}
\includegraphics[angle=-90,width=7cm]{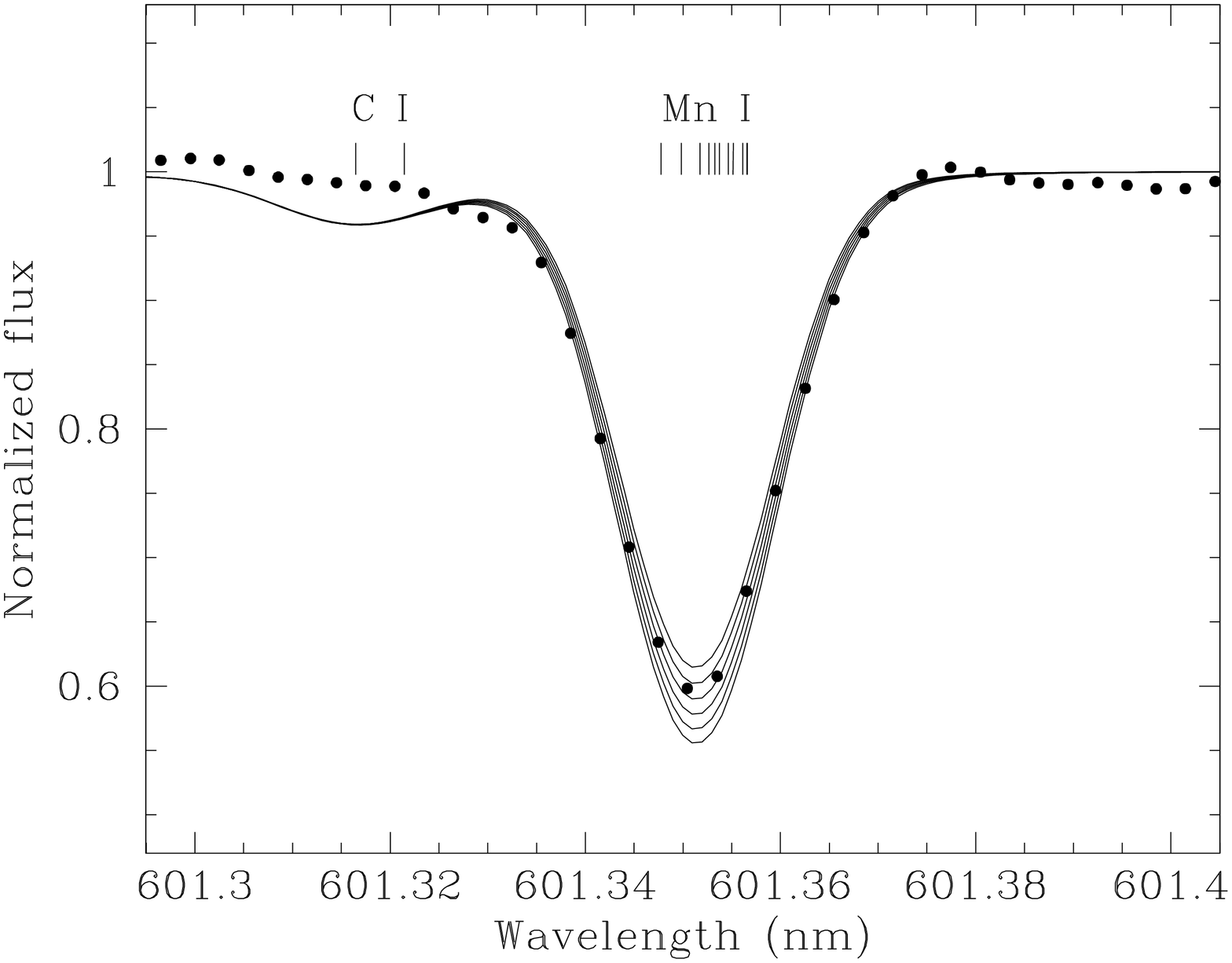}
\includegraphics[angle=-90,width=7cm]{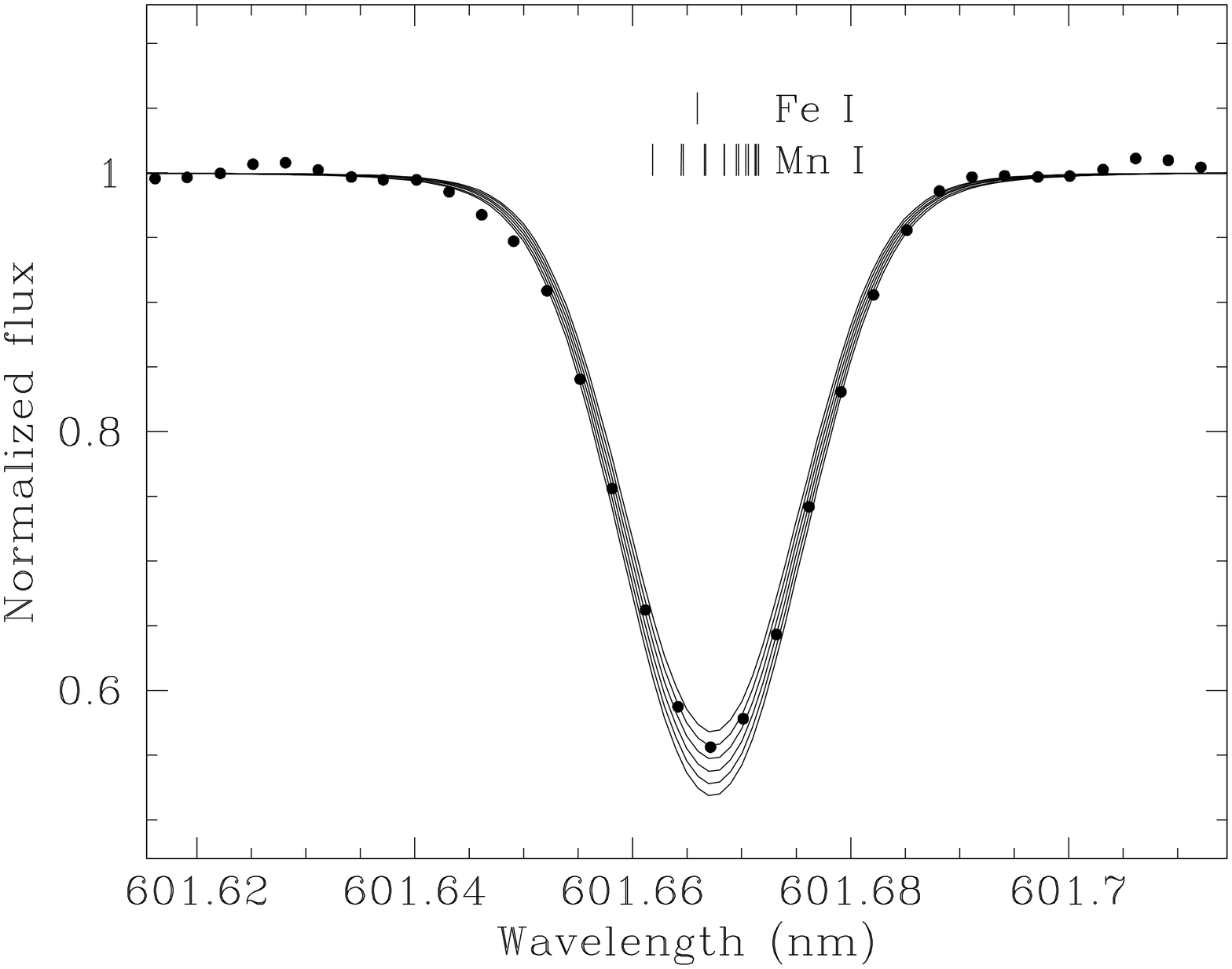}
\caption{Mn\,{\sc i} lines used in the final abundance analysis. Here 
we show the solar spectrum with $\bullet$ and five synthetic
spectra used to derive the final abundances for
 each of them. For each line we also show the components
of the hyper fine structure and the position of some adjacent lines.
In two cases we also include a blending Fe\,{\sc i} in  the analysis.}
\label{lines.fig}
\end{figure*}

The synthetic spectra for the Mn\,{\sc i} lines were first convolved
with a Gaussian line profile to compensate for the instrumental
broadening and after that they were convolved with a rad-tan profile
to account for the broadening due to the combined effects of rotation
and macro turbulence (see Sect.\,\ref{rotmacro.sect}).  After the
continuum was set a $\sim \chi^2$ was calculated in the same way as
when deriving the $v_{\rm r+m}$ and the $\sim \chi^2$ was minimized to
estimate the abundance.  An example for how the Mn abundance was
derived for the Mn\,{\sc i} line at 539.4\,nm is shown in
Fig.\,\ref{ab.fig}.

Finally, all the derived abundances were normalized, line by line, to
the solar values that we obtained from our own spectra of scattered
solar light (Sect.\,\ref{sect:stars}).  This means that the Sun will
always have [Mn/H] $=0$.  It is the mean value of these normalized
values that we are using in the Sect.\,\ref{sect:res} and
\ref{sect:interp} and which are the basis for the [Mn/H] values quoted
in Table\ref{stars.tab}.

\subsection{Errors}
\label{sect:errors}

\begin{table}
\caption{Effects on derived Mn abundances due to errors in the stellar
parameters. For four stars we show the effect on the resulting abundance,
line by line (columns two to five), 
when  $T_{\rm eff}$ is increased by 70 K, when $\log g$ is increased
by 0.1 dex, and when [Fe/H] is increased by 0.1 dex. }
\label{errors.tab}
\centering
{\begin{tabular}{l c  c  c  c  c  c  c}
\hline \hline
 \rule{0ex}{2.5ex} &\multicolumn{4}{c}{$\varepsilon$\,(Mn)} \\
& 539.4 & 543.2 & 601.3 & 601.6 &&  \\
& [nm] & [nm] & [nm] & [nm] \\
\hline
{\bf{Sun}} with FEROS &  5.35 & 5.30 & 5.21 & 5.36\rule{0pt}{2ex}  \\
 with SOFIN &5.22 & 5.19 & -- & --\\
\hline
{\bf{HIP3142}}               & 4.82 & 4.85 & 4.69 & 4.80 \rule{0pt}{2ex}  \\
$\Delta$ $T_{\rm eff}=+70$ K & 4.89 & 4.92 & 4.73 & 4.84 \\
$\Delta \log g = +0.1$       & 4.82 & 4.84 & 4.68 & 4.80 \\
$\Delta$ [Fe/H]$=+0.1$       & 4.85 & 4.87 & 4.72 & 4.80 \\
\hline
{\textbf{HIP88622}}          & 4.64 & 4.68 & 4.63 & 4.74\rule{0pt}{2ex} \\
$\Delta$ $T_{\rm eff}=+70$ K & 4.72 & 4.77 & 4.67 & 4.78 \\
$\Delta \log g = +0.1$       & 4.64 & 4.67 & 4.62 & 4.73 \\
$\Delta$ [Fe/H]$=+0.1$       & 4.68 & 4.70 & 4.65 & 4.75 \\
\hline
{\textbf{HIP103682}}         & 5.63 & 5.62 & 5.58 & 5.77 \rule{0pt}{2ex } \\
$\Delta$ $T_{\rm eff}=+70$ K & 5.72 & 5.71 & 5.63 & 5.83 \\
$\Delta \log g = +0.1$       & 5.63 & 5.62 & 5.59 & 5.77 \\
$\Delta$ [Fe/H]$=+0.1$       & 5.65 & 5.65 & 5.60 & 5.80 \\
\hline
{\textbf{HIP118115}}         & 5.23 & 5.21 & 5.07 & 5.27 \rule{0pt}{2ex} \\
$\Delta$ $T_{\rm eff}=+70$ K & 5.33 & 5.30 & 5.13 & 5.33 \\
$\Delta \log g = +0.1$       & 5.23 & 5.21 & 5.07 & 5.27 \\
$\Delta$ [Fe/H]$=+0.1$       & 5.26 & 5.25 & 5.10 & 5.30 \\
\hline                  
\end{tabular}}
\end{table}

In this work we are concerned with a differential study. This means
that the Mn abundance derived from each Mn\,{\sc i} line is normalized
to the abundance we derived from our own solar spectra for that
particular line (see Sect.\,\ref{sect:derab}).  Furthermore, we are
studying stars that span a fairly narrow range in $T_{\rm eff}$ as
well as in $\log g$ (see e.g.  Table\,\ref{stars.tab}). This means
that when we compare two stars at a given [Fe/H] then any differences
in the derived Mn are, to first order, independent of the assumptions
made for the stellar parameters and in the modeling of the line.

There are two types of errors that we in general are concerned with --
statistical and systematic. We will here quantify both of them,
however, bearing in mind that for the interpretation of our results in
Sect.\,\ref{sect:interp} it is mainly the statistical errors that
concerns us.

\paragraph{Line-to-line scatter}

\begin{figure}
\centering
\includegraphics[width=8cm]{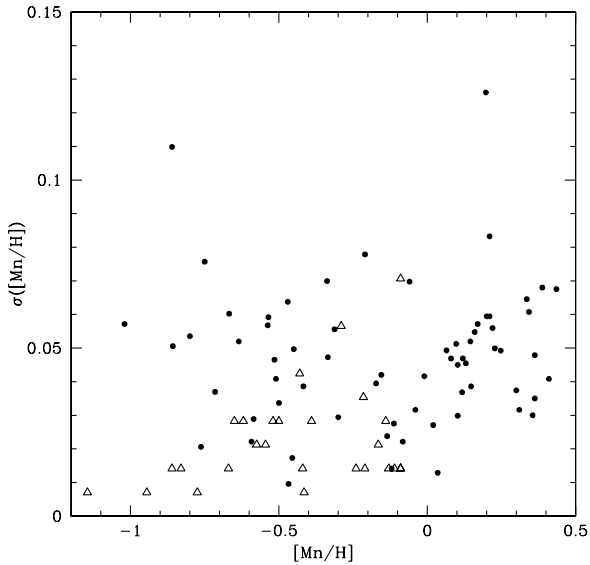}
\caption{$\sigma$([Mn/H]) vs. [Mn/H] for all stars. $\bullet$ denote
  stars observed with FEROS and $\triangle$ stars observed with
  SOFIN. For all stars observed with FEROS four lines were used whilst
  for SOFIN only two (see Sect.\,\ref{sect:stars}).  For the FEROS
  observations the mean error is 0.047 dex in [Mn/H].  }
\label{sigma.fig}
\end{figure}

In Fig.\,\ref{sigma.fig} we show the $\sigma$([Mn/H]) arising from the
line-to-line scatter for each star using the values normalized to the
solar values (see Sect\,\ref{sect:sellines}).  As can be seen the
resulting $\sigma$ are small, generally below 0.06 dex with a mean
error for the FEROS observations of 0.047. This gives a formal error
in the mean\footnote{The formal error in the mean is defined as
  $\sigma/\sqrt {N_{\rm lines}}$ where $\sigma$ is the normal standard
  deviation and ${N_{\rm lines}}$ are the number of lines used in the
  abundance determination (see e.g. Gray 1992).}  of 0.024 dex.  For
stars observed with SOFIN only two lines are used and hence the
$\sigma$ are of less value to estimate the internal
consistency. However, for most stars observed with SOFIN the two lines
are in good agreement. Furthermore, there is no sign
that the scatter increases with increasing Mn abundance (and hence
increasing Fe abundance).  In summary we find that the internal
statistical errors due to line-to-line scatter are small and an easily
quantifiable source of scatter in any abundance trend.

\paragraph{The $\chi^2$ estimation of the abundances}
We have used an un-normalised $\chi^2$ to estimate the final
abundances rather than a $\chi$-by-eye procedure. Our procedure would
give a best value also when e.g. the hfs in the lines are not well
reproduced in the line-synthesis (i.e. due to incomplete or erroneous
atomic data). However, since we always do additional visual
inspections of the fits we we believe that for the data set presented
in this work this type of systematic error is rather small which is
evidenced in e.g. Figs.\,\ref{ab.fig} and \ref{sigma.fig}.  We have
also extensively checked our derived abundances, line-by-line, as a
function of the stellar parameters and found no obvious trends (see
Sect.\,\ref{sect:nlte}).  This together with the fact that the
internal scatter between Mn abundances derived from different lines is
small (see paragraph above and Fig.\,\ref{sigma.fig}) implies that the
hfs, as well as other aspects of the shape of the lines have been
taken care of.

\paragraph{Setting of continuum}
The exact placement of the continuum is sometimes
difficult. Especially so when there are many lines close to the line
that is being analyzed which results in an overall depression of the
apparent continuum. Figure\,\ref{lines.fig} shows the spectrum in the
vicinity of each of the four Mn\,{\sc i} lines used in this study. As
can be seen these, small, regions of the spectra would (in most cases)
not be enough to judge the continuum level wherefore we relied on a
larger wavelength range to set the continua. In general the continua
are harder to set in the most metal-rich stars as the lines are
getting stronger in such stars.  However, this appears to be of no
great concern in our study as the line-to-line scatter does in fact
not increase when [Mn/H], and thus also [Fe/H], increases.  We note
that from the SOFIN spectra of the Sun, for lines at 539.4 and
543.2\,nm, we derive somewhat lower Mn abundances than from the FEROS
spectra (see Table\,\ref{errors.tab}). This could be due to that we
placed the continua too low in the SOFIN spectra. However, we have
not been able to quantify this.  As we normalize our final abundances
to that of our own solar abundance, line-by-line, this will not be a
concern for our final conclusions.

\paragraph{Erroneous hyper fine structure (hfs)}
Prochaska \& McWilliam\,(2000) showed that the hfs data used by Nissen
et al.\,(2000) gave erroneous abundances. Figure\,2 in Prochaska \&
McWilliam\,(2000) shows the effect of applying the correct hfs to the
data by Nissen et al.\,(2000). In particular they found that [Mn/Fe]
increased by $\sim$0.05 dex and that the effect was larger below
[Fe/H] $\sim -0.4$. Recently, del Peloso et al.\,(2005) have compared
different sets of hfs. In particular they compare the hfs from
R. Kurucz's electronically available linelists\footnote{ available at
  {\tt http://kurucz.harvard.edu/linelists.html}} and
Steffen\,(1985). Nissen et al.\,(2000) used the data by
Steffen\,(1985).  In good agreement with Prochaska \&
McWilliam\,(2000), del\,Peloso et al.\,(2005) find that the two sets
of hfs result in Mn abundances that show increasingly different
[Mn/H] as [Fe/H] decreases and that the differences increases
progressively from [Fe/H]$\sim -0.4$ in the sense that the data from
Steffen\,(1985) give the smaller [Mn/H]. As discussed in
Sect.\,\ref{sect:sellines} we use the hfs (both as concerns wavelength
as well as the strength of the different hfs components) listed by
Prochaska et al.\,(2000) in their Table 19\footnote{Note - in the
  electronic version of Prochaska et al.\,(2000) the electronically
  readable versions of the tables have been switched around so that
  Table 18 is in fact Table 19 (in the machine readable versions of
  the tables).}. In our analysis we have payed special attention to
how well the structure of each Mn\,{\sc i} line is fitted. Through
inspections of the fitting of the line at 602.1\,nm in $\sim$15 stars
with a large span in [Fe/H] we came to the conclusion that the hfs for
this particular line as listed in Prochaska et al.\,(2000) (which is
based on R.\,Kurucz's electronically available hfs lists) is not a
good representation of the actual shape of the line. The miss-match
was particularly obvious in stars with small values for the combined
rotation and macro turbulence broadening, i.e. where lines are
sharper. This line was hence discarded from further analysis.  For the
other lines we found no such discrepancies. We thus believe that, at
the accuracy of our observed spectra, the hfs we use is a good
representation of the actual line profiles.

\begin{figure*}
\centering
\includegraphics[angle=0,width=8cm]{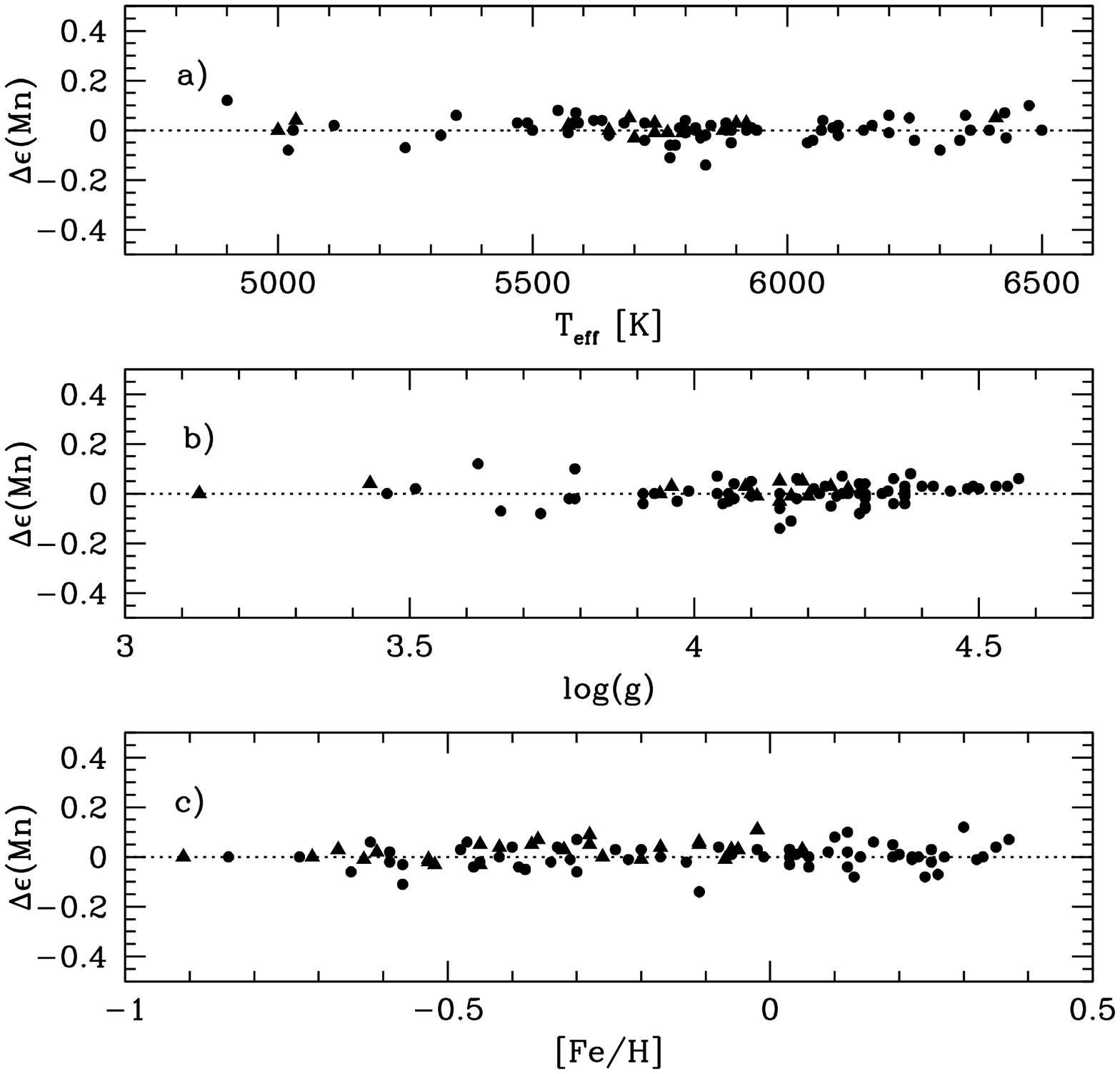}
\includegraphics[angle=0,width=8cm]{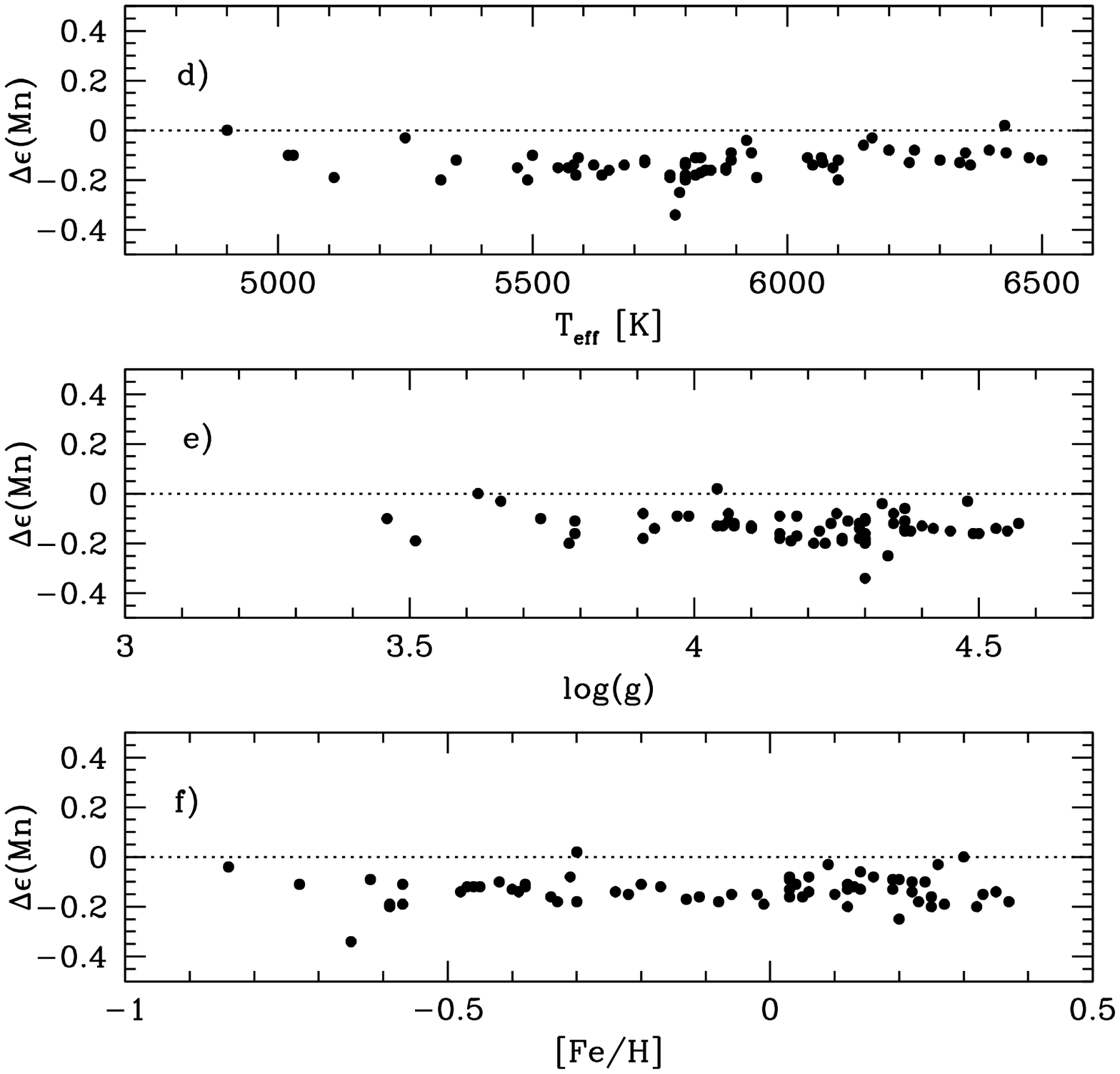}
\caption{Panels {\bf a -- c} show the differences between the Mn
  abundance derived from the Mn\,{\sc i} lines at 539.4 and 543.2\,nm
  as a function of $T_{\rm eff}$, $\log g$, and [Fe/H].  The mean
  offset between the Mn abundance derived from these two lines is, for
  stars observed with FEROS, 0.002$\pm$0.047 dex and for stars
  observed with SOFIN it is 0.026$\pm$0.035 dex. Stars observed with
  FEROS are marked with $\bullet$ and stars observed with SOFIN with
  $\blacktriangle$.  Panels {\bf d -- f} show the differences between
  the Mn abundance derived from the Mn\,{\sc i} lines at 601.3 and
  601.6\,nm as a function of $T_{\rm eff}$, $\log g$, and [Fe/H].  The
  mean offset between the Mn abundance derived from these two lines is
  --0.133 $\pm 0.057$ dex.  }
\label{difflines.fig}
\end{figure*}

\paragraph{Erroneous stellar parameters}
In Table\,\ref{errors.tab} we summarize our standard error analysis
where we have varied the relevant stellar parameters one by one and
then performed the abundance analysis in exactly the same manner as
done in the main analysis. As we have taken all stellar parameters
from the work by Bensby et al.\,(2003 \& 2005) we use their estimates
of the uncertainties in the stellar parameters.  We thus vary $T_{\rm
  eff}$ by +70 K, $\log g$ by +0.1 dex, and [Fe/H] by +0.1 dex. As is
clear from Table\,\ref{errors.tab} all of these errors individually
result in errors on the resulting $\varepsilon$(Mn) of less than 0.1
dex. The largest error by far comes from an erroneous $T_{\rm eff}$
while an error in $\log g$ is utterly negligible, and an error in
[Fe/H] gives a smaller error on the resulting Mn abundance. We may
thus assume that errors in the adopted stellar parameters contribute
no more than $\sim 0.07$ dex to the overall error in [Mn/H].

\subsubsection{Deviations from non-LTE?}
\label{sect:nlte}

Our current analysis does not include a full non-LTE analysis of the
line formation for the Mn\,{\sc i} lines. However, in order to exclude
that our results should be compromised by this omission we have done
extensive checks to exclude that departures from LTE is a major
concern for the stars that we study.

The lines we kept for our final analysis divide into two pairs. One
set of lines with a lower excitation energy of 0.0 eV and one with a
lower excitation energy of 3.07 eV (see Table\,\ref{mnlines.tab}).  We
started by investigating these two line pairs separately.

The differences in derived Mn abundance from the lines at 539.4\,nm
and 549.2\,nm show no trends with $T_{\rm eff}$, $\log g$, or
[Fe/H]. Furthermore, the differences between the two lines are small
with a low scatter (see Eq.\,\ref{eq1} and \ref{eq3}) (Figs.
\,\ref{difflines.fig} a--c).

For the line pair at $\lambda=$ 601.3 and 601.6\,nm the comparison,
again, showed no obvious trends with either $T_{\rm eff}$, $\log g$,
or [Fe/H] (Fig.\,\ref{difflines.fig} d--e, note that there are no
SOFIN observations for these lines).  However, there is a significant
offset between the Mn abundance derived from the two lines
(Eq.\,\ref{eq2}).  Both of these two lines also show offset with
respect to the abundances derived from the lines at 539.4 and
543.2\,nm. The abundance derived from the line at 601.3\,nm is offset
by --0.074 dex from the mean Mn abundance derived from the two lines
at 539.4 and 543.2\,nm, while the abundance derived from the line at
601.6\,nm is offset by +0.059 dex. The standard deviation is in both
cases 0.06 dex.

\begin{figure}
\centering
\includegraphics[angle=0,width=8cm]{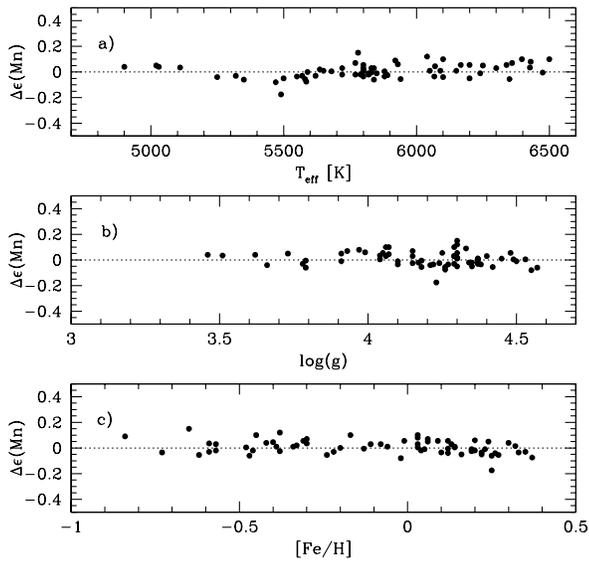}
\caption{Differences between the mean Mn abundances derived from the
  Mn\,{\sc i} lines at 539.4 and 543.2\,nm and the two lines at 601.3
  and 601.6\,nm as a function of $T_{\rm eff}$, $\log g$, and [Fe/H].
  The mean offset between the Mn abundance derived from these two line
  pairs is 0.007 dex with a $\sigma$ of 0.055 dex.  }
\label{difflines3.fig}
\end{figure}

We find no trends with either [Fe/H] or $\log g$ for the difference
between the mean of the 539.4\,nm and 543.2\,nm and the mean of the
601.3\,nm and 601.6\,nm lines (see Figs\,\ref{difflines3.fig}b.  and c).

When we do the same comparison but now as a function of $T_{\rm eff}$
we do find what might be a pattern, Fig\,\ref{difflines3.fig}a., such
that there appears to be a ``dip'' in the difference at $\sim 5400$ K,
spanning perhaps $\pm 250$ K. For stars hotter than $5750$ K there
appears to be hardly any trend. Also for stars with temperatures below
$\sim 5100$ K the trend appears flat as well, although here we are
dealing with only a few stars so the conclusion will be less robust.
Note, that also when we compare the two lines in multiplet 27 one by one
to the mean value derived from the two lines of multiplet 1 the same
pattern shows up.

These comparisons show that the excitation balance, most probably, is
not influenced by departure from LTE. However, departures from
ionizational balance may still be present.

\paragraph{Absolute abundances}\label{prov}
For the Sun, as observed with FEROS, we derive $\varepsilon$\,(Mn) =
5.35$\pm$0.05 and 5.30 $\pm 0.05$ dex for the lines at 539.4 and
543.2\,nm, respectively. Asplund et al.\,(2005) list a value for the
solar photosphere of $5.39\pm 0.03$ dex for the Sun. Taking possible
error sources into account, especially setting of continua and
treatment of hfs, we conclude that the $\varepsilon$\,(Mn) we derive
for the Sun, observed with the FEROS spectrograph, from the the 
lines from Multiplet 1 agree very well with the standard value.

For our observation of the Sun with SOFIN we derive $5.27 \pm 0.05$
and $5.26 \pm 0.06$ for the lines at 539.4 and 543.2\,nm,
respectively.  The internal consistency is also here good, in fact
better than for the FEROS spectra. However, the absolute values
deviate more from the standard value than the abundances based on
the FEROS spectra.

For the lines at 601.3\,nm and 601.6\,nm (multiplet 27)
we derive $\varepsilon$\,(Mn) = 5.21 $\pm$ 0.06 and 5.36 $\pm$
0.05 dex, respectively.  Again, the value derived from the line
at 601.6\,nm agrees very well with the standard value.  Could the line
at 601.3\,nm suffer from departures from LTE? If this was the case it
would in general be likely that the line that suffered from non-LTE
effects should show changes as a function of e.g. $T_{\rm eff}$ or
another atmosphere parameter. In this case we see no such thing, only
an offset from the other line in the same multiplet and, to first
order, one would attribute the offset to an erroneous log$gf$-value
for the line at 601.3\,nm.

In summary, it appears that the three lines at 539.4\,nm, 543.2\,nm,
and 601.6\,nm all yield Mn abundances that are, within the errors, in
accordance with the standard value for the solar photosphere, whilst
the line at 601.3\,nm gives $\varepsilon$\,(Mn) about 0.15 dex lower
than the other lines. A first interpretation of this deviation is that
the offset is due to the errors in the derivation of the
log$gf$-value.

\paragraph{Chromospheric activity?}
It has been suggested that the strength of the Mn\,{\sc i} lines at
539.4 and 543.2\,nm would be a measure of chromospheric activity in
the Sun and similar stars (Livingston \& Wallace 1987; Booth et
al.\,1984). In fact, Livingston \& Wallace (1987) found that the
Mn\,{\sc i} line at 539.4\,nm varied in strength, both central
intensity and equivalent width, in lock step with variations of the
Ca\,{\sc ii}\,K line at 393.3\,nm which is sensitive to chromospheric
activity. Danilovi\'c et al.\,(2005) re-analysed the Kitt Peak data,
which is un-evenly distributed in time, and confirmed that the
Mn\,{\sc i} line changes with three periods; 11 years, quasi-biannual,
and a 27-day period.

However, the size of the changes in equivalent width are
small. Moreover, although we find a difference between the abundances
derived for the Sun from the spectra recorded by the FEROS and SOFIN
spectrographs, which are taken about two years apart, once the
data for all stars have been normalized to the solar value
(i.e. putting the Sun at 0 dex) there is no difference between the
data recorded with the two different spectrographs.  This indicates
that the size of the variations induced by the chromospheric activity,
if real, are sufficiently small that it does not effect elemental
abundance analysis at the level of resolution and S/N that our data
have.

\subsubsection{Summary}

In summary we find that the statistical errors in our derived
abundances are small.  The line-to-line scatter is less than 0.06 dex
and the mean error in the mean is 0.047 dex. Statistical errors in the
stellar parameters also give low errors, on the same order as the
line-to-line scatter.

Systematic errors are more difficult to assess. We find one $\log
gf$-value that might be erroneous. Further measurements of the $\log
gf$-values for these lines are desirable.

Errors arising from the limitations in the modeling include
e.g. non-LTE effects and neglect of chromospheric activity. Both are
possibilities for the lines we use. From an empirical inspection of
our results we find no basis for claiming that such effects are
present in the spectra we have analysed.

\section{Results}
\label{sect:res}

\begin{table}
  \caption{Mn abundances line by line for stars observed with FEROS. 
    The wavelengths of the lines used are indicated in the header row. 
The last column indicates if the star is classified as thin or
 thick disk star (see Bensby et al. 2003 \& 2005). }             
\label{abferos.tab}      
\centering          
\begin{tabular}{lllllllllllllllllll l l l llllllll}     
\hline\hline       
ID &  539.4 & 543.2 &  601.3 &  601.6 & Disk\\	
  & [nm] & [nm] & [nm] & [nm] \\            			    
\hline 
Sun	      & 5.35 &  5.30 &   5.21 &  5.36  &   \\
HIP3086	     &  5.04 &  5.18 &   5.00 &  5.16  &   Thick disk \\
HIP3142	     & 4.83 &   4.85 &   4.68 &  4.80  &   Thin disk\\
 ... & ... & ... & ... &  ... &  ... \\
 ... & ... & ... & ... &  ... &  ... \\
 ... & ... & ... & ... &  ... &  ... \\
\hline                  
\end{tabular}
\end{table}

\begin{table}
\caption{Mn abundances line by line for stars observed with SOFIN. 
    The wavelengths of the lines used are indicated in the header row. 
The last column indicates if the star is classified as thin or
 thick disk star (see Bensby et al. 2003 \& 2005).
 }             
\label{absofin.tab}      
\centering          
\begin{tabular}{llllllll            l l l llllllll}     
\hline\hline
ID &  539.4 & 543.2 & Disk\\
  & [nm] & [nm] &  \\            			    
\hline                    
Sun	   & 5.27 & 5.26	 & \\	
HIP699	   & 5.02 & 5.03	 & Thin disk\\
HIP910	   & 4.87 & 4.80	 & Thin disk\\
... & ... & ... & ... \\
... & ... & ... & ... \\
... & ... & ... & ... \\ 
\hline                  
\end{tabular}
\end{table}

\begin{figure}
\centering
\includegraphics[angle=-0,width=9cm]{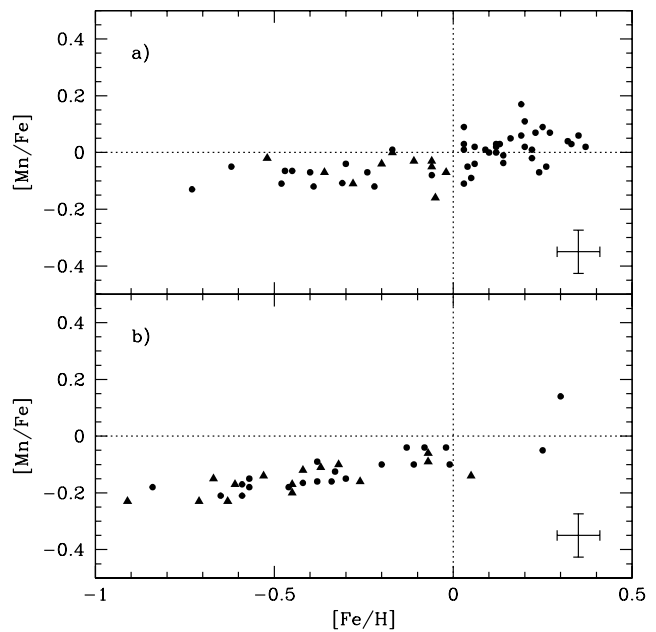}
\caption{ {\bf a.} [Mn/Fe] vs [Fe/H] trends for the sample with
  kinematics typical of the thin disk.  $\bullet$ indicate stars
  observed with FEROS and $\blacktriangle$ indicate stars observed
  with SOFIN.  {\bf b.} [Mn/Fe] vs [Fe/H] trends for the sample with
  kinematics typical of the thick disk.  $\bullet$ indicate stars
  observed with FEROS and $\blacktriangle$ indicate stars observed
  with SOFIN.  Representative error-bars are indicated in the lower
  right hand corners.  }
\label{mnfedisks.fig}
\end{figure}

\begin{figure}
\centering
\includegraphics[angle=-0,width=9cm]{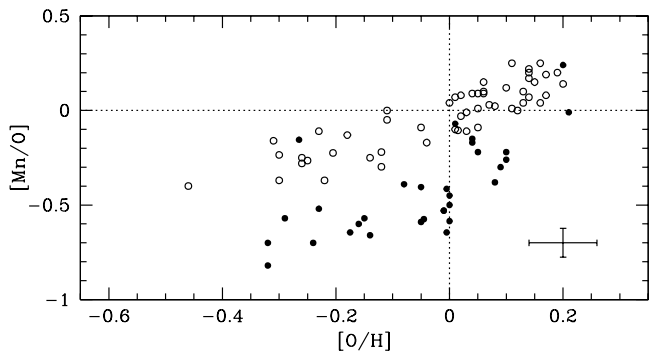}
\caption{[Mn/O] as a function of [O/H]. The oxygen data are taken from
  Bensby et al.\,(2004) and Bensby et al.\,(2005). Stars with
  kinematics typical of the thick disk are shown as filled circles and
  the stars with kinematics typical of the thin disk are shown as open
  symbols. Typical error-bars are shown in the lower right hand
  corner.  }
\label{oxygen.fig}
\end{figure}

The abundance results for all our stars are summarized in
Tables\,\ref{stars.tab}, \ref{abferos.tab}, and \ref{absofin.tab} and
shown in Fig.\,\ref{mnfedisks.fig}.  For stars with kinematics typical
of the thick disk we see a steady increase in the [Mn/Fe] as a
function of [Fe/H], while for the stars with kinematics typical of the
thin disk we see an essentially flat trend below [Fe/H] = 0 and a
shallow increase at higher metallicities.

We also combine our new Mn abundances with oxygen abundances from
Bensby et al.\,(2004 \& 2005).  In Fig.\,\ref{oxygen.fig} the trend of
[Mn/O] vs. [O/H] is shown. For the stars with kinematics typical of
the thin disk we simply see a steady increase in [Mn/O] as [O/H]
increases.

Stars with kinematics typical of the thick disk show a similar trend,
albeit offset, for [O/H] $<$ 0. For higher [O/H] there is a hint of a
faster increase in [Mn/O]. However, based on the current data set this
must remain a tentative observation.

In summary we find that when plotting [Mn/O] vs. [O/H] the two samples
show trends with similar slopes but an offset of about 0.3 dex in
[Mn/O].

\paragraph{Comparison with Reddy et al.\,(2006)}
We note that Reddy et al.\,(2006) find an [Mn/Fe] vs. [Fe/H] trend for
their thick disk sample that is exactly similar to what they found for
their thin disk sample (Reddy et al.\,2003).  This is in contradiction
with our results as shown in Fig.\,\ref{mnfedisks.fig}. We have no
simple explanation to this difference between the two
investigations. A possible explanation lies in the exact lines used
and the methods. Reddy et al.\,(2006) use the line at 602.1 nm.  This
line we found to be poorly reproduced by available hfs linelists (see
Sect.\,\ref{sect:errors} and paragraph ``Erroneous hyper fine
structure (hfs)''). Furthermore, Reddy et al.\,(2006), although they
also synthesise the lines instead of comparing directly with the
stellar spectra themselves they compare only the measured equivalent
width with the equivalent with produced in the line
synthesis. Although in principle an excellent approach it could be
possible to get an incorrect answer if the hfs is poorly reproduced
(compare discussion in Sect.\,\ref{sect:errors} paragraph ``The
$\chi^2$ estimation of the abundances'').  However, we do not have
enough information to claim that such an error has occurred.

\section{Interpretation of abundance trends}
\label{sect:interp}

That Mn is made in explosive nucleosynthesis appears well established
through theoretical modeling, e.g. Arnett\,(1996). However, it is
still uncertain how much Mn is expelled from the SNe, i.e. the yield,
and it also remains uncertain which type of SN are the main
contributor to the Mn enrichment, i.e. core collapse or degenerate
systems. It is also controversial how metallicity dependent the yields
may or may not be (see e.g. McWilliam et al. 2003 and Carretta et
al. 2004 for a discussion of the observational evidence).

McWilliam et al.\,(2003) extensively discussed the available data for
Mn and concluded that it is most likely made in SN\,II and that the
yields are metallicity dependent. Furthermore, from the data for the
Sgr dSph they found preliminary evidence that Mn may also be made in
SN\,Ia. Carretta et al.\,(2004) used Mn as an example to explore the
origins of elements with odd atomic numbers.  By including two Sgr
dSph stars analysed by Bonifacio et al.\,(2000) they concluded that
the differences between the galactic field stars discussed in
McWilliam et al.\,(2003) and their Sgr dSph RGB stars were less
compelling and thus that the case for significant production of Mn in
SN\,Ia was weakened.  Our new data for two kinematically selected
samples significantly increases the information available for Mn at
disk-like metallicities. We will therefore revisit the data discussed
in McWilliam et al.\,(2003) and Carretta et al.\,(2004) and combine
that with our new data as well as data for halo stars in the Milky Way
to infer the origin of Mn.

\subsection{A selection effect}

McWilliam et al.\,(2003) revisited the Mn data from Nissen et
al.\,(2000).  In particular they applied a correction to the Nissen et
al.\,(2000) Mn values to correct for an erroneous hyper fine structure
that had been used in that study.  In McWilliam et al.\,(2003) they
noted that for the disk sample they saw a jump in [Mn/Fe]. This jump
appear at [Fe/H]$\sim -0.7$. McWilliam et al.\,(2003) question if this
apparent jump is real.  Our data shows that the jump can be explained
as a result of the selection process in Nissen et al.\,(2000).  Our
study consists of two stellar samples: one with kinematics that are
typical for the thick disk and one which has kinematics typical of the
thin disk in our Galaxy. Our stars are confined to the metallicity
range which is typical of the Galactic disks, i.e. with an [Fe/H] of
$-1$ to $+0.5$ dex. The sample of stars in Nissen et al.\,(2000) are
from the list by Chen et al.\,(2000). In Chen et al.\,(2000) the stars
were selected to represent the thin and the thick disks, respectively.
However, the criteria they used are such that they selected
exclusively thin disk stars for [Fe/H]$>$--0.7 and only thick disk
stars for [Fe/H]$<$--0.7 (see also discussion in Bensby et al.\,2003
and their Fig.\,14 ).  If we from our sample,
Fig.\,\ref{mnfedisks.fig}a, take only thick disk stars for
[Fe/H]$<$--0.7 and only thin disk stars for [Fe/H]$>$--0.7 then the
resulting abundance trend would be very similar to the one in Nissen
et al.\,(2000) as seen in the corrected abundances from Prochaska \&
McWilliam (2000).  We thus conclude that the jump in [Mn/Fe] vs [Fe/H]
found in previous studies is due to a sample selection effect and not
an intrinsic property of the disks. It is, however, worth noting that
the Nissen et al.\,(2000) indeed is the first study that indicated
that the Mn trends for the thin and the thick disks do differ. A
finding that we confirm and extend to a much larger [Fe/H] range, in
particular for the thick disk.

\subsection{Comparison with oxygen}
\label{sect:oxygen}

\begin{figure}
\centering
\includegraphics[angle=-0,width=9cm]{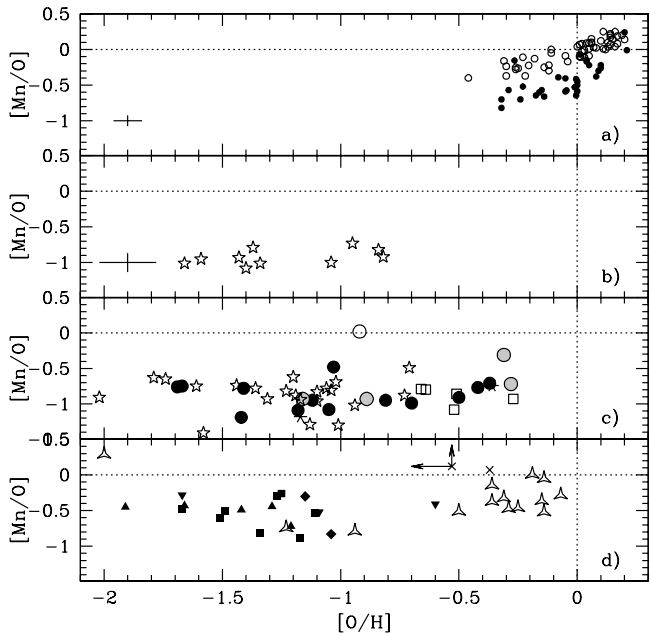}
\caption{Mn and O abundances for several samples collected from the
  literature.  {\bf a.} Our own samples (compare
  Fig.\,\ref{oxygen.fig}). $\bullet$ shows the stars with kinematics
  typical of the thick disk and $\circ$ stars with kinematics typical
  of the thin disk. {\bf b.} Stars from the study by Bai et
  al.\,(2004).  {\bf c.} Mn abundances from Sobeck et al.\,(2006) and
  O abundances from Fulbright \& Johnson (2003). Stars that have
  kinematics typical of the thin disk are shown as $\circ$, stars that
  have kinematics typical of the thick disk as $\bullet$, and stars
  with kinematics typical of the halo as open stars. Transition objects between the
  thin and thick disk are shown as grey circles (for the definition of
  these classifications see Bensby et al.\,2003). Object that are not 
classifiable and falls between then halo and thick disk distributions
are marked by $\ast$. See also Fig.\ref{toomre.fig}.
 The stars are a
  mix of dwarfs stars, sub-giants, and giants (compare
  Fig.\,\ref{hr.fig}). Five stars with Mn from Sobeck et al.\,(2006)
  and O from Mel\'endez et al.\,(2006) are shown as $\square$.  {\bf
    d.}  Mn and O abundances for bright red giant stars in dwarf
  spheroidal galaxies.''Squashed'' triangles indicate the stars in the Sgr dSph
  (McWilliam \& Smecker-Hane 2005).  $\blacksquare$ indicate stars in
  Sculptor from Geisler et al.\,(2005) and Shetrone et al.(2003). We
  also include data for LeoII, Fornax and, Carina from Shetrone et
  al.\,(2003). They are shown as $\Diamondblack$, $\blacktriangle$,
  and a filled upside down triangle, respectively.  Two stars from
  Bonifacio et al.\,(2000) are shown as $\times$. For one of them O
  does only have an upper limit. This is indicated by arrows.  }
\label{individual.fig}
\end{figure}

\begin{figure}
\centering
\includegraphics[angle=-0,width=9cm]{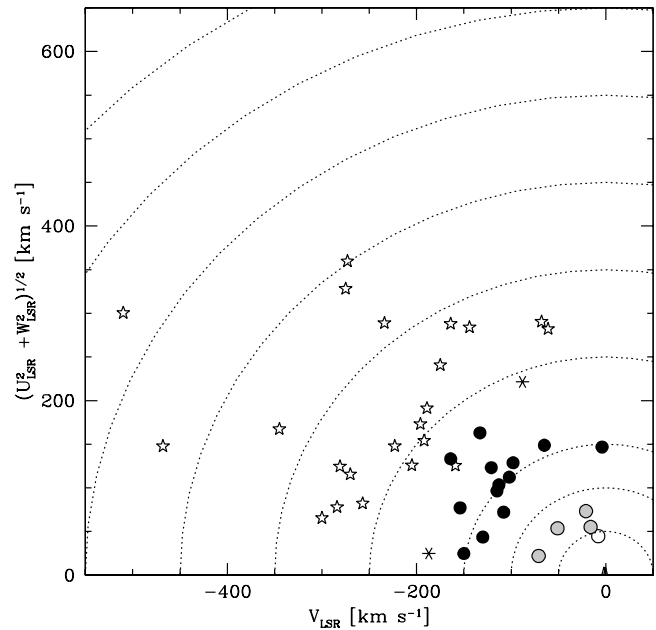}
\caption{Toomre diagram showing the kinematics for the stars with Mn
  abundances from Sobeck et al.\,(2006). Stars with kinematics typical
  of the thin disk are shown as $\circ$, stars with kinematics typical
  of the thick disk as $\bullet$, and transition objects between the
  thin and thick disk are shown as grey circles (for the definition of
  these classifications see Bensby et al.\,2003). Stars with typical
  halo kinematics are shown as open stars while transition objects
  between the halo and thick disk are shown as $\ast$.  }
\label{toomre.fig}
\end{figure}

\begin{figure}
\centering
\includegraphics[angle=-0,width=9cm]{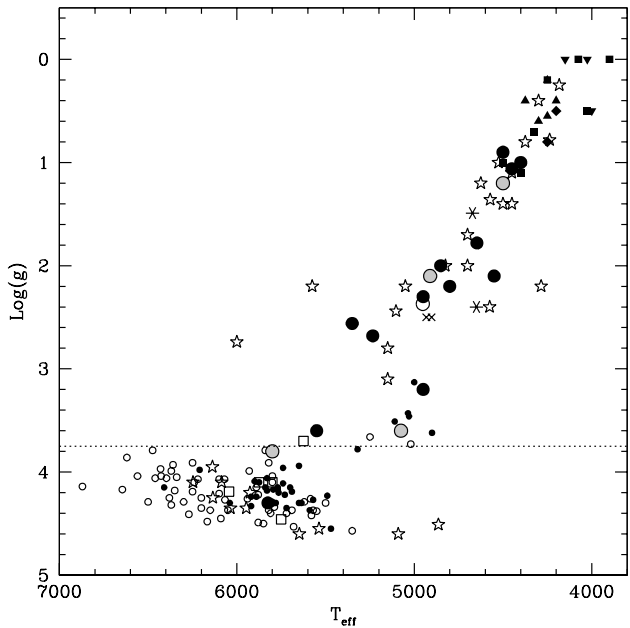}
\caption{$\log g$ - $T_{\rm eff}$ diagram for all the stars shown in
  Fig.\,\ref{individual.fig} and \ref{mnomw.fig}. Symbols are the same
  as in Fig.\ref{individual.fig} where in particular $\circ$
  represents stars with kinematics typical of the thin disk and
  $\bullet$ stars with kinematics typical of the thick disk, whilst
  stars with halo kinematics are shown as open stars.  RGB stars from
  dSphs are shown with a range of symbols as identified in
  Fig.\ref{individual.fig}.  }
\label{hr.fig}
\end{figure}

An $\alpha$-element, like oxygen, is a better reference element
than Fe if we want to investigate which supernovae an element (mainly)
comes from. The main reason for this is that oxygen is only made in
SN\,II whilst Fe is made in both SN\,II and in SN\,Ia (e.g. Timmes et
al. 1995). With O as the reference element the interpretation of any
abundance trend hence becomes simpler.  Oxygen is preferred over
$\alpha$-elements such as Ca and Mg since those elements have a small
contribution from SN\,Ia (Timmes et al.\,1995).

The main reason why previous studies have not used O extensively as the
reference element is because it is difficult to derive O abundances
from stellar spectra for three reasons: the O lines are few and hence
not always present in the available stellar spectra; the stronger O
lines (the triplet at 777 nm) have significant non-LTE effects (see
e.g. Kiselman 1993); and the lines that do not suffer from non-LTE
effects are very weak, in fact so weak that they may not be observable
at all in more metal-poor stars.

Over the last years an extensive discussion has taken place in the
literature about the exact [O/Fe] vs [Fe/H] trend at metallicities
below [Fe/H] $\leq -1$ (see for example volume 45 issue 8 of New
Astronomy and discussions in Nissen et al.\,2002 and Asplund et
al.\,2004).  However, with the advent of larger telescopes and a
better understanding of how to compensate for the non-LTE effects that
are present in the triplet lines the number of stars with reliable O
abundances have increased enormously.

In order to study the origin of Mn, i.e. which type of supernovae are
the main contributors and if the yields are metallicity dependent, we
have collected a representative set of studies of stars with Mn as
well as O data available in the literature. In particular we have
selected 42 dwarf and giant stars from the new study of Mn by Sobeck
et al.\,(2006) [Mn data kindly provided by the authour prior to
publication] which have O abundances in Fulbright \& Johnson (2003);
14 red giant stars in the Sgr dSph from the study by McWilliam \&
Smecker-Hane (2005) [data kindly provided by the authours].  In
addition to these we have included additional data for stars in dSph
galaxies as well as for halo dwarf stars.  Below we briefly comment on
the different studies and how many stars we have selected from each.

\paragraph{Sobeck et al.\,(2006) $+$ Fulbright \& Johnson (2003)} 
Sobeck et al.\,(2006) provide Mn abundances for 214 dwarf and giant
stars with kinematics typical of (mainly) the (metal-poor) thick disk
and the halo. Of these we have selected 42 stars with O abundances in
Fulbright \& Johnson (2003). We use the O abundances derived from the
forbidden oxygen line at 630.0 nm using, the recommended, Alonso
temperature scale. The reader is referred to Fulbright \& Johnson
(2003) for an in-depth discussion of the different temperature scales,
their shortcomings, and why the Alonso scale should be more trusted
than the other two scales investigated. The abundances are shown in
Fig.\,\ref{individual.fig}d.  Kinematic data for the stars were taken
from Fulbright (2002) whenever possible. For stars with no kinematic
data in Fulbright (2002) we used data from Simmerer et al.\,(2004).
The kinematic data were used to calculate probabilities that the stars
belong to the thin disk, thick disk, or the halo, respectively, using
the method described in Bensby et al.\,(2003 \& 2005).  
In Fig.\ref{toomre.fig} we show the Toomre diagram for
these stars.

\paragraph{Bai et al.\,(2004)} A study of 10 metal-poor halo dwarf
stars. Equivalent widths are used in the abundance
determination. Oxygen is derived from the permitted triplet lines and
corrected for non-LTE effects according to Takeda (2003).  The
abundance data are shown in Fig.\,\ref{individual.fig}b.

\paragraph{Melendez et al.\,(2006)} From this study of 31 turn-off
stars in the range of [Fe/H] --3.2 to --0.7 dex we have taken oxygen
data for five stars for which Sobeck et al.\,(2006) have obtained Mn
abundances for. These stars are placed in a particularly interesting
range of [O/H] and partly bridges the gap between halo and thick disk
(Fig.\,\ref{individual.fig}c) as well as providing confirmation of the
trends indicated by the more numerous RGB data from Sobeck et
al.\,(2006).

\paragraph{McWilliam \& Smecker-Hane (2005) $+$ McWilliam et
  al.\,(2003)} These studies of stars in the Sgr dSph galaxy
gives O and Mn abundances for 14 RGB stars based on high-resolution
spectra. The analysis is based on measured equivalent widths but hfs
has been taken into account when the elemental abundances were
calculated.  The abundance data are shown in
Fig.\,\ref{individual.fig}d.

\paragraph{Geisler et al.\,(2005) $+$ Shetrone et al.\,(2003)} Oxygen
and Mn abundances for four red giants in the Sculptor dwarf spheroidal
galaxy (Geisler et al. 2005) are combined with data for another four
Sculptor RGB stars from Shetrone et al.\,(2003).  We have normalized
the O and Mn abundances for the two data-sets.  From Shetrone et
al.\,(2003) we also take data for two stars in the LeoII dSph, five
stars in the Carina dSph, and three stars in the Fornax dSph.  The
abundance data are shown in Fig.\,\ref{individual.fig}d.

\paragraph{Bonifacio et al.\,(2000)} In this study spectra for two RGB
stars in the Sgr dSph were analysed. The analysis is based on measured
equivalent widths and hfs is not taken into account. For completeness
we do, however, include them in our discussion (see
Sect.\,\ref{sect:intro}).  The abundance data are shown in
Fig.\,\ref{individual.fig}d.

\begin{figure}
\centering
\includegraphics[angle=0,width=9cm]{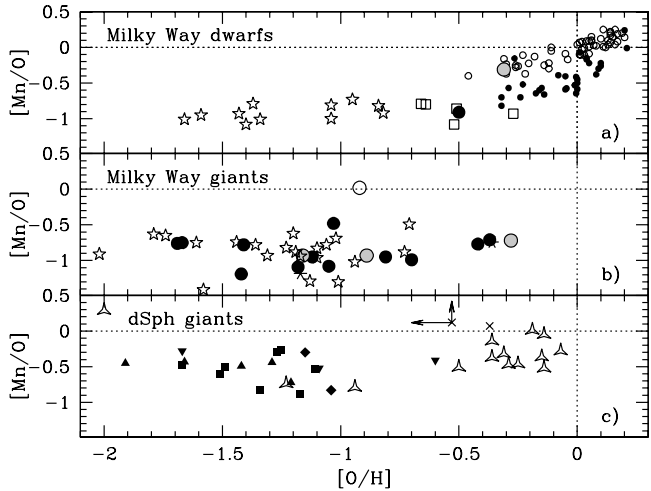}
\caption{{\bf a.} Shows the [Mn/O] vs [O/H] trend for dwarf stars in
  the two disks and the halo. $\bullet$ indicate thick disk stars and
  $\circ$ thin disk stars from this study. The one large $\bullet$ is
  from Sobeck et al.\,(2006) as are the transition objects marked with
  grey circles and the halo stars from Bai et al.\,(2004) are marked
  with open stars.  $\square$ marks stars with O from Mel\'endez et
  al.\,(2006) and Mn from Sobeck et al.\,(2006).  {\bf b.} [Mn/O] vs
  [O/H] trend for evolved stars in the Milky Way.  $\bullet$ indicate
  thick disk stars, $\circ$ thin disk stars, grey circles transition
  objects, and open stars mark the halo stars.  All stars have Mn from Sobeck et
  al.\,(2006). {\bf c.} Stars in other galaxies.  Stars in Sculptor
  dSph are indicated by $\blacksquare$ (from Shetrone et al.\,2003 and
  Geisler et al.\,2005). Stars in LeoII, Carina, and, Fornax are
  marked by $\Diamondblack$, $\blacktriangle$, and a filled upside
  down triangle, respectively. All data from Shetrone et
  al.\,(2003). Two stars from Bonifacio et al.\,(2000) are shown as
  $\times$. For one of them O does only have an upper limit. This is
  indicated by arrows.  Giants in the Sgr dSph are shown as open,
  ``squashed'' triangles.}
\label{evolstages.fig}
\end{figure}

In Fig.\,\ref{hr.fig} we show the stellar parameters ($T_{\rm eff}$
and $\log g$) for all of the stars described above. As can be seen the
data span a large range in $\log g$.  In Fig.\,\ref{evolstages.fig} we
have divided the stars according to their evolutionary status to see
if any, obvious, effects/trends in the abundance ratios are due to
stellar evolution.  We used $\log g$ to divide the stars into dwarfs
and giants. An arbitrary cut was imposed at 3.75 dex (compare
Fig.\,\ref{hr.fig}). From Fig.\,\ref{evolstages.fig} we conclude that
although the dwarf star samples show much tighter and better defined
trends the more evolved stars overall follow the same trends.

Figure\,\ref{evolstages.fig}c shows that the RGB stars in
the dSph galaxies are enhanced in [Mn/O] relative to the Milky Way
giants in Fig.\,\ref{evolstages.fig}b.  For example the dSph RGB stars
from all galaxies studied by Shetrone et al.\,(2003) have $<$[Mn/O]$>
=-0.54 \pm 0.19$ and the RGB stars in the Sgr dSph have $<$[Mn/O]$>
=-0.34 \pm 0.29$ (McWilliam et al.\,2005) while the Milky Way halo
giants from the compilation of data from Sobeck et al.\,(2006) (Mn
data) and Fulbright \& Johnson\,(2003) (oxygen data) have $<$[Mn/O]$>
=-0.87 \pm 0.23$.

We note that the giant stars studied by Sobeck et al.\,(2006) that
have $\log g < 1.0$ do not show enhanced [Mn/O] relative to the less
evolved stars in their sample. In fact, the stars with $\log g < 1.0$
are some of the stars with the lowest [Mn/O] in that sample.  Hence,
it does appear that the dSph stars show an abundance pattern that is
enhanced in relation to what we see for similar stars in the Milky Way
and that the enhancement, probably, should be sought in environmental
effects (e.g. a slower star formation history) rather than being a
result of stellar evolution.  However, to establish if the effect is
real further studies of halo and dSph RGB stars should be
undertaken. These studies should be done in a differential manner to,
to first order, be able to exclude modeling errors.

We note that one star with thin disk kinematics (Mn data from Sobeck
et al.\,2006) have an unusually high [Mn/O]
(Fig.\,\ref{individual.fig}c). We have no explanation for this.

\begin{figure*}
\centering
\includegraphics[angle=0,width=14cm]{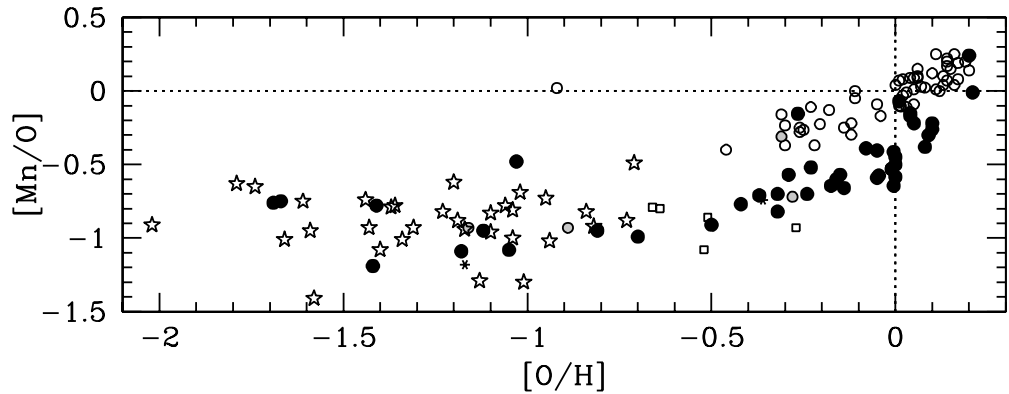}
\caption{[Mn/O] vs [O/H] for the Milky Way. The data are the same as
  in Figs.\,\ref{individual.fig}a, \ref{individual.fig}b, and
  \ref{individual.fig}c and Figs.\,\ref{evolstages.fig}a and \ref{evolstages.fig}b but here we distinguish the different
  kinematic components instead of distinguishing the different
  studies.  $\circ$ indicates stars with kinematics typical for the
  thin disk, $\bullet$ indicate stars with kinematics typical of the
  thick disk, grey circles indicate transition objects between the
  thin and thick disk, open stars indicate stars with kinematics typical
  of the halo, and $\ast$ transition objects between the thick disk
  and the halo. $\square$ shows the five stars with O abundances from
  Mel\'endez et al.\,(2006). No kinematic information is available for
  these stars.  }
\label{mnomw.fig}
\end{figure*}

\subsection{The origin of Mn}

We use the final compilation of data for Milky Way stars in
Fig.\,\ref{mnomw.fig} to investigate the origin of Mn in the Milky
Way. For the halo and metal-poor thick disk, [O/H]$\leq -0.5$, the
[Mn/O] trend is flat.  This indicates that the production of Mn and O
are well balanced. Moreover, we know from the study of Bensby et
al.\,(2004) that the archetypal signature of SN\,Ias in the thick disk
do not occur until [O/H] = 0. Hence the up-going trend we see after
[O/H] $\simeq -0.5$ must be interpreted as being due to metallicity
dependent Mn yields in SN\,II. The rising trend seen for the thin disk
sample could also be interpreted in this fashion. Although here we do
know that SN\,Ia contribute to the chemical enrichment and hence the
increase might also be due to these objects.

The dSph galaxies provide an interesting comparison. As discussed in
Sect.\,\ref{sect:oxygen}, the RGB stars in the dSph appear to be
genuinely more enhanced in [Mn/O] at a give [O/H] than the Milky Way
halo and thick disk giants. The higher [Mn/O] values for the RGB stars
in the dSph could be interpreted as being evidence for a slower star
formation in those galaxies as compared to the star formation in our
own halo and thick disk. For the Sgr dSph we note that the [Mn/O]
vs. [O/H] trend does appear to resemble the trend for the thick disk
(our data plus data from the literature and as shown in
Fig.\,\ref{evolstages.fig}). Perhaps indicating that the Sgr dSph has
a more Milky Way like star formation history than the other dSph. But,
these are tentative conclusions and need further
confirmation based on larger, and differential, abundance studies.

\section{Summary}
\label{sect:sum}

We have analysed four Mn\,{\sc i} lines in 95 dwarf stars previously
studied by Bensby et al.\,(2003 \& 2005). The stars were selected to
have kinematics typical of the thick or the thin disk. Using these two
well defined and well studied stellar samples we find that the
abundance trends in the two samples differ such that the stars with
thin disk kinematics are enhanced in Mn relative to stars with thick
disk kinematics.

We also find that the previously reported ``step'' in the [Mn/Fe] vs
[Fe/H] trend for disk stars in the Milky Way (see e.g. McWilliam et
al.\,2003) is in fact an artifact due to incomplete sampling of the
two disk populations. Thus there is no need to invoke a large spread
in the age-metallicity relation for the thin disk needed to explain
such a ``step'' (compare discussion in McWilliam et al.\,2003).

Furthermore, when comparing the Mn abundances with Fe abundances the
thick disk stars show a steadily increasing trend of [Mn/Fe]
vs. [Fe/H] whilst the stars with kinematics typical for the thin disk
show a flat trend up and until [Fe/H] = 0 and after that an increasing
trend.

In order to further study the origin of Mn we have combined our new Mn
abundances with O abundances. Fe is made both in SN\,II and in
SN\,Ia. By using O, which is only made in SN\,II, as the reference
element we simplify the interpretation of the abundance data. For our
stars we took the O abundances from Bensby et al.\,(2004) and added
data from a number of other studies of (mainly) giant stars in the
disks and halo of the Milky Way as well as giant stars in dSph
galaxies.

Our interpretation is that these data, to first order, can be
explained by metallicity dependent yields in SN\,II. This is,
essentially, in agreement with the conclusions in McWilliam et
al.\,(2003).

However, it is not possible from these data to exclude that SN\,Ia
contribute to the Mn production at least in the thin disk. 
Observations that would be helpful to further clarify the nature
of Mn includes:

\begin{enumerate}
\item higher quality Mn determinations for stars in the metal-poor
  regime (today the scatter is large enough that it could hide a true
  trend)
\item increased number of stars, with known kinematics, with [O/H]
  around --0.5 dex (Fig.\,\ref{mnomw.fig} shows a lack of stars in
  this regime )
\item larger samples of halo and metal-poor thick disks stars,
  preferably dwarf stars, analysed in a differential study to
  definitely rule out any differences between these two components
\end{enumerate}

We are currently working on data-sets that will provide steps towards
these goals.

\begin{acknowledgements}
  We would like to thank Bengt Gustafsson, Bengt Edvardsson, Kjell
  Eriksson, and Martin Asplund for usage of the MARCS model atmosphere
  program and their suite of stellar abundance (EQWIDTH) and synthetic
  spectrum generating programs. Jennifer Sobeck and co-authours are
  thanked for providing their Mn data prior to publication. Tammy
  Smecker-Hane and Andy McWilliam are thanked for sharing their Mn and
  O data for the Sgr dSph.  SF is a Royal Swedish Academy of Sciences
  Research Fellow supported by a grant from the Knut and Alice
  Wallenberg Foundation. TB acknowledges support from the National
  Science Foundation, grant AST-0448900.  This work has made use of
  the SIMBAD database operated at CDS, Strasbourg, France, and the
  VALD database at {\tt http://www.astro.uu.se/}$\sim${\tt vald\/}.
\end{acknowledgements}


\begin{thebibliography}{}

\bibitem[]{} Arnett, W.D., 1971, ApJ, 166, 153

\bibitem[]{} Arnett, W.D., 1996, \emph{Supernovae and nucleosynthesis}, Princeton series in astrophysics

\bibitem[]{} Asplund, M., Gustafsson, B., Kiselman, D., \& Eriksson, K., 1997, A\&A, 323, 286

\bibitem[]{} Asplund, M., Grevesse, N., Sauval, A.J., Allende Prieto, C., \& Kiselman, D., 2004, A\&A, 417, 751

\bibitem[]{} Asplund, M., Grevesse, N., \& Sauval, A.J., 2005, 
\emph{Cosmic Abundances as Records of Stellar Evolution and Nucleosynthesis in honor of David L. Lambert}, page 25, Eds, Barnes, T.G. III \& Bash, F.N.


\bibitem[]{} Bai, G.S., Zhao, G., Chen, Y.Q., Shi, J.R., Klochkova, V.G., Panchuk, V.E., Qiu, H.M., 
\& Zhang, H.W., 2004, A\&A, 425, 671

\bibitem[]{} Barklem, P.S., Piskunov, N., \& O'Mara, B.J., 2000, A\&AS, 142, 467

\bibitem[]{} Bensby, T., \& Feltzing, S., 2006, MNRAS, 367, 1181

\bibitem[]{} Bensby, T., Feltzing, S., \& Lundstr\"om, I., 2003, A\&A, 410, 527

\bibitem[]{} Bensby, T., Feltzing, S., \& Lundstr\"om, I., 2004, A\&A, 415, 155

\bibitem[]{} Bensby, T., Feltzing, S., Lundstr\"om, I., \& Ilyin, I., 2005, A\&A, 433, 185

\bibitem[]{} Bonifacio, P., Hill, V., Molaro, P., et al., 2000, A\&A, 359, 663

\bibitem[]{} Booth, A.J., Blackwell, D.E., Petford, A.D., \& Shallis, M.J., 1984, MNRAS, 208, 147

\bibitem[]{} Carretta, E., Gratton, R.G., Bragaglia, A., Bonifacio, P., \& Pasquini, L., 2004,
A\&A, 416, 925

\bibitem[]{} Chen, Y. Q., Nissen, P. E., Zhao, G., Zhang, H. W., \& Benoni, T., 2000, A\&AS, 141, 491

\bibitem[]{} Chieffi, A. \& Limongi, M., 2004, ApJ, 608, 405 

\bibitem[]{} Danilovi\'c, S., Vince, I., Vitas, N., \& Jovanovi\'c, P., 2005, Serb. Astron. Journal, 170, 79

\bibitem[]{} del Peloso, E.F., Cunha, K., da Silva, L., \& Porto de Mello, G.F., 2005, A\&A, 441, 1149

\bibitem[]{}Edvardsson, B., Andersen, J., Gustafsson, B., Lambert, D.L.,
Nissen, P.E., \& Tomkin, J., 1993, A\&A 275, 101

\bibitem[]{} Fulbright, J.P., 2002, AJ, 123, 404

\bibitem[]{} Fulbright, J.P. \& Johnson, J.A.,  2003, ApJ, 595, 1154

\bibitem[]{} Geisler, D., Smith, V.V., Wallertstein, G., Gonzalez, G., \& Charbonnel, C.
2005, AJ, 129, 1428


\bibitem[]{} Gratton, R.G., 1989, A\&A, 208, 171

\bibitem[]{} Gray, D.F., 1992, 
\emph{'The observation and analysis of stellar photospheres'}, Cambridge University Press, Cambridge

\bibitem[]{} Gustafsson, B., Bell, R.A., Eriksson, K., \& Nordlund,  
{\AA}., 1975, A\&A 42, 407

\bibitem[]{} Holweger, H., Heise, C., \& Kock, M., 1990, A\&A, 232, 510 

\bibitem[]{} Kiselman, D. 1993, A\&A, 275, 269

\bibitem[]{}  Kupka, F., Piskunov, N.E., Ryabchikova, T.A., Stempels, H.C., \&
Weiss, W.W., 1999, A\&AS, 138, 119

\bibitem[]{} Limongi, M., Chieffi, A., 2005, ASPC, 342, 122

\bibitem[]{} Livingston, W., \& Wallace, L., 1987, ApJ, 314, 808 

\bibitem[]{} McWilliam, A., \& Smecker-Hane, T.A., 2005, ASP Conf. Ser., 336, 221

\bibitem[]{} M\"ackle, R., Holweger, H., Griffin, R., \& Griffin, R., 1975, A\&A, 38, 239

\bibitem[]{} McWilliam, A., Rich, R.M., \& Smecker-Hane, T.A., 2003, ApJ, 592, L21

\bibitem[]{} Mel\'endez, J., Shchukina, N.G., Vasiljeva, I.E., \& Ram\'irez, I., 2006, ApJ, 642, 1082

\bibitem[]{} Nissen, P.E., Chen, Y.Q., Schuster, W.J., \& Zhao, G., 2000, A\&A, 353, 722

\bibitem[]{} Nissen, P.E., Primas, F., Asplund, M., \& Lambert, D.L.,  2002 ,A\&A, 390, 235

\bibitem[]{} Piskunov, N.E., Kupka, F., Ryabchikova, T.A., Weiss, W.W., \&
Jeffery, C.S., 1995, A\&AS, 112, 525

\bibitem[]{}  Prochaska, J.X., \& McWilliam, A., 2000, ApJ, 537, L57

\bibitem[]{}  Prochaska, J.X., Naumov, S.O., Carney, B.W., McWilliam, A., \& Wolfe, A.M., 2000, AJ, 120, 2513

\bibitem[]{} Reddy, B.E., Lambert, D.L., \& Allende Prieto, C., 2006, MNRAS, 367, 1329

\bibitem[]{} Reddy, B.E., Tomkin, J., Lambert, D.L., \& Allende Prieto, C., 2003, MNRAS, 340, 304

\bibitem[]{} Ryabchikova, T.A. Piskunov, N.E., Stempels, H.C., 
Kupka, F., \& Weiss, W.W. 1999, Proc. of the 6th International 
Colloquium on Atomic Spectra and Oscillator Strengths, Victoria BC, 
Canada, 1998, Physica Scripta, T83, 162

\bibitem[]{} Shetrone, M., Venn, K.A., Tolstoy, E., Primas, F., Hill, V.,
\& Kaufer, A., 2003, AJ, 125, 684

\bibitem[]{} Simmerer, J., Sneden, C., Cowan, J.J., Collier, J., Woolf, V.M., Lawler, J.E.,
2004, ApJ, 617, 1091

\bibitem[]{} Simmons, G. J., \& Blackwell, D. E., 1982, A\&A, 112, 209

\bibitem[]{} Sobeck, J.S., Ivans, I.I., Simmerer, J.A., Sneden, S., Hoeflich, P.,
Fulbright, J.P., \& Kraft, R., 2006, AJ, 131, 2949

\bibitem[]{} Steffen, M., 1985, A\&AS, 59, 403

\bibitem[]{} Takeda, Y., 2003, A\&A, 402, 343

\bibitem[]{} Timmes, F.X., Woosley, S.E., \& Weaver, T.A., 1995, ApJS, 98, 617

\bibitem[]{} Woosley, S.E. \& Weaver, T.A., 1995, ApJS, 101, 181
\end{thebibliography}
\end{document}